\documentclass{aa}  

\usepackage{graphicx}
\usepackage{txfonts}
\usepackage{lipsum}
\usepackage{subcaption}         
\usepackage{lscape}             
\usepackage{placeins}           

\begin{document} 

   \title{Deimos photometric properties: analysis of 20 years of observations (2004-2024) by the Mars Express HRSC camera}

   \author{A. Wargnier\inst{1,2}
          \and
          P. N. Simon\inst{3}
          \and
          S. Fornasier\inst{1}
          \and
          N. El-Bez-Sebastien\inst{1}
          \and
          D. Tirsch\inst{5}
          \and
          K.-D. Matz\inst{5}
          \and
          T. Gautier\inst{2,1}
          \and
          A. Doressoundiram\inst{1}
          \and
          M. A. Barucci\inst{1}
          }

   \institute{
            LIRA, Observatoire de Paris, Université PSL, Sorbonne Université, Université Paris Cité, CNRS, CY Cergy Paris Université, 5 place Jules Janssen, Meudon, 92195, France\\
            \email{antonin.wargnier@obspm.fr}
         \and
            LATMOS/IPSL, CNRS, Université Versailles St-Quentin, Université Paris-Saclay, Sorbonne Université, 11 Bvd d’Alembert, Guyancourt, F-78280, France
         \and
             Laboratoire d'Astrophysique de Marseille (LAM), France
         \and
            Institute of Space Science, German Aerospace Center (DLR), Rutherfordstrasse 2, 12489 Berlin, Germany 
             }

   \date{Received 19 May 2025; accepted 5 September 2025}
 
  \abstract
   {}
   {The goal of this study is to analyze the photometric properties of Deimos using observations obtained by the Mars Express (MEX) mission, aiming at improving the photometric properties and providing new insights into the texture and composition of the surface of the smallest Martian moon. The findings also support the Martian Moon eXploration mission (MMX) observations.}
   {We analyzed the data obtained by the High Resolution Stereo Camera (HRSC) and the Super Resolution Channel (SRC) onboard MEX. The HRSC data, obtained through the use of four filters (blue, green, red, and IR), have a spatial resolution ranging from 390 to 800 m/px. In comparison, the SRC panchromatic data have a resolution ranging from 85 to 300 m/px. The SRC data are of particular interest due to their coverage of a wide range of phase angles, including the opposition effect of Deimos (0.06-138°). Observations of both HRSC and SRC cover only the Mars-facing side of Deimos. As the SRC camera was never absolutely calibrated before and during the MEX mission, we performed the first absolute calibration of the SRC using observations of Jupiter and stars. We then performed the disk-integrated and disk-resolved photometric analysis using the Hapke model.}
   {The Deimos surface is dark and predominantly backscattering. The single-scattering albedo (SSA) value (between 6.8\% and 7.5\%, depending on the model) is similar to the one derived from Phobos. The Deimos phase curve shows a strong opposition effect due to shadow-hiding, with almost no effect of the coherent-backscattering process. The amplitude and the half-width of the shadow-hiding opposition surge were found to be 2.14 $\pm$ 0.14 and 0.065 $\pm$ 0.004, respectively. We found a very high porosity of 86\% at the top-layer surface ($\sim10$ µm), consistent with the tentative presence of complex-shaped grains or fractal aggregates. Such a high porosity would likely also indicate the presence of a thick dust layer. We did not observe significant variations of the opposition surge across the surface. We observed a blue unit on Deimos in a similar way to Phobos, located on the streamers, themselves on the equatorial ridge. The Deimos blue unit exhibits variations relative to its average surface that are similar to those of the blue unit on Phobos, characterized by an average reflectance increase of about 35\% (and up to 58\%) and a spectral slope decrease of 50\%. This blue unit may be due to a different texture of the surface between the two units, with finer grain and/or higher porosity. In contrast to the "blue unit" photometric behavior exhibited by Phobos on several crater rims, no such behavior has been observed for Deimos.}
   {The Deimos photometric properties, including the SSA, opposition surge, and phase integral, are very similar to Phobos. The presence of a blue unit on Deimos reinforces the idea that the Martian moons have a common origin. The capture of two different bodies with similar spectroscopic and photometric properties appears very unlikely.}


   \titlerunning{Deimos photometric properties from Mars Express HRSC observations}
   \authorrunning{A. Wargnier et al.}

   \maketitle

\section{Introduction}
Deimos, the smallest Martian moon, is relatively unknown, orbiting at a distance of 23,000 km further than Phobos, the innermost and largest of the Martian moons. If the Martian moons share important similarities such as albedo, VNIR spectrum, and density, Deimos exhibits distinct characteristics when compared to Phobos. Previous observations by spacecraft orbiting Mars have revealed a more homogeneous surface on Deimos, with the absence of the grooves that are a characteristic geological feature of Phobos. The Mars near-side of Deimos exhibits several significant craters, in particular two craters named Voltaire and Swift with a diameter of about 1.9 km and 1 km, respectively. Voltaire is the largest crater on the surface of Deimos and was created 134 Ma ago by an impactor having a diameter $>25$ meters \citep{Nayak_2016}. Despite the presence of craters, the surface of Deimos is relatively smooth in comparison with Phobos, which may be due to sesquinary impact gardening \citep{Nayak_2016}. Deimos also exhibits a particular shape compared to Phobos, with an important equatorial ridge and a large concavity of 11 km at the southern pole corresponding to almost two times the mean radius. The origin of such geological features is not established, but it could be related to the accretion of large blocks during the (re)formation of Deimos or to a large impact cratering process that occured on the Martian moon. This last hypothesis may be more probable as Deimos shows the presence of resurfacing processes by ejecta \citep{Thomas_1989}. The remarkably smooth appearance of Deimos' surface may be attributed to the significant impact that resulted in the formation of an ejecta blanket -- estimated to be a few hundred meters thick -- which was subsequently influenced by seismic activity following the impact event \citep{Thomas_1989, Thomas_1996}. \par
Previous photometric investigations of Deimos were performed by \cite{Pang_1983}, \cite{French_1988}, \cite{Thomas_1996}, and more recently by \cite{Fraeman_2012}. Using Mariner 9 data, \cite{Noland_1977} showed that the surface of Deimos exhibits bright patches and regions (in particular related to the ridge ("streamers") and to the crater rims; \citealt{Thomas_1980}). \cite{French_1988} found that the differences between bright and dark regions on Deimos' surface may be simply related to particle size effect, with smaller grains ($<$40 µm) responsible for the higher reflectance. The extensive study of Deimos' photometric properties by \cite{Thomas_1996} showed that the opposition effect for Deimos is smaller than for Phobos and that Deimos is brighter (20-30\%) than Phobos for phase angles larger than 10 degrees. The bright and dark regions on Deimos exhibit no variations in phase curve, which is not in agreement with the hypothesis of different grain sizes by \cite{French_1988} that should produce different phase curve behavior. If the craters' rims exhibit generally brighter materials than the interior, \cite{Thomas_1996} noticed that few craters on Deimos have also darker excavated materials in their surroundings. \par
The High-Resolution Stereo Camera (HRSC) and its Super Resolution Channel (SRC) onboard Mars Express (MEX) provide a unique opportunity, at present, to constrain the photometric properties. Despite the polar and close-to-Mars orbit limiting the observation to the Deimos Mars side, the quantity of data acquired during 20 years with different illumination conditions, in particular covering the opposition effect, is particularly important to study in detail the photometric properties of Deimos surface. We therefore analyzed the images obtained by the HRSC and SRC cameras to provide new insights into the surface of Deimos. This work will also be useful in support of the JAXA/Martian Moon eXploration (MMX) mission operation planning. The recent flyby of Deimos by the Hera spacecraft will also provide important information on this Martian moon.

\section{Observations and data reduction}
The HRSC and SRC data were retrieved from the ESA Planetary Science Archive and the HRSC team. We selected HRSC data available in the four absolutely calibrated blue (BL), green (GR), red (RE), and infrared (IR) filters. \par
We used HRSC data of level 3, which corresponds to radiometrically calibrated data, including removal of the contribution of dark and flat field images, and computation of the absolute calibration factors. We also used level 3 SRC data. The calibration process of the SRC images includes correction of the dark current and the dark signal uniformity, division by the flat field images, and the removal of hot pixels. However, no absolute calibration processes were performed prior to the launch, or are available so far. In this work, we propose an attempt to absolutely calibrate the SRC camera using images containing stars and Jupiter in the field-of-view.\par
\begin{figure}
\centering
  \resizebox{\hsize}{!}{\includegraphics{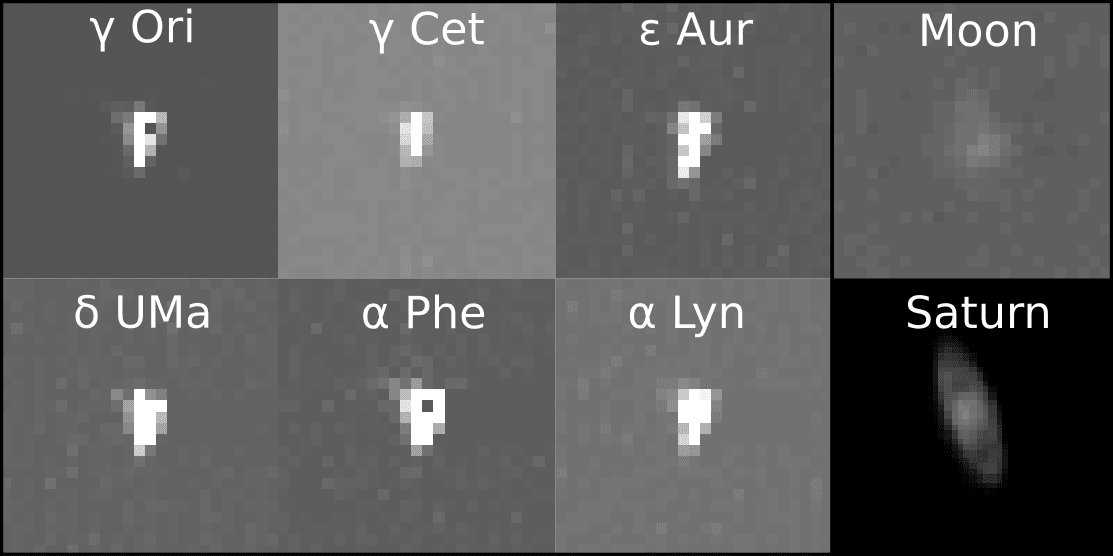}}
  \caption{Example of SRC mutual events observations -- from the Martian orbit -- of different types of objects: stars, Saturn, and the Moon. The shape of the stars is not symmetric because of the astigmatism observed in the SRC point-spread function (PSF, \citealt{Oberst_2008}). The $\gamma$ Ori and $\alpha$ Phe images are examples of saturated stars. Saturated images were discarded for the calibration.}
  \label{calib_SRC_objects}
\end{figure}

Looking at non-resolved objects (e.g., stars) in the SRC images, therefore at the instrument point-spread function (PSF), it appears that this PSF has a very unusual non-symmetric shape. According to \cite{Oberst_2008}, this is due to the fact that the camera suffered from important astigmatism (see Fig. \ref{calib_SRC_objects}). Our investigation also revealed slight modifications of the full width at half maximum (FWHM) from star to star. We did not perform specific deconvolution to remove the effects of the point-spread function (PSF), as our primary focus was to measure the flux. \par
We also noticed that the SRC images are affected by the presence of unexpected pixel values on the right of a bright object, such as Deimos. This would be likely due to an electronic effect, whereby a capacitor may have exhausted its charge due to the high signal, and the sudden drop in light intensity may have resulted in some undervoltage until the capacitor was full again. The values of the pixel in this small region are set to zero, and therefore do not significantly affect the computation of the flux from the target.

\subsection{HRSC data analysis}
Due to the distance between MEX and Deimos, only a few observations using HRSC were acquired. We collected 18 images in the four different calibrated filters (blue, green, red, and IR). These images were acquired between 2018-01-07 and 2024-11-25 with a spatial resolution ranging from 390 m/px to almost 800 m/px, and a phase angle from 1° to 90°. For each of the filters, one image (2018-10-17) was systematically removed because of an acquisition issue. \newline
The HRSC data in the four filters were converted to radiance factor (i.e., I/F) using the correction factors provided in the header of each image. The I/F calibration factor takes into account the specific spectral behavior of each filter and the heliocentric distance of the observations. The performed analysis on the HRSC Deimos images is similar to the method described in \cite{Fornasier_2024} for Phobos HRSC data: (i) SPICE simulations were performed using the latest MEX SPICE kernels and the latest shape model from \cite{Ernst_2023}. As HRSC images for Deimos have a size of about 5100 $\times$ 600 pixels, and as Deimos is only 20 $\times$ 20 pixels in the center of the images, the simulation was performed using the mean time of the acquisition. After the simulation procedure, we obtained a set of six images: original, incidence, emission, phase, latitude, and longitude. (ii) We coregistered the SRC image with the simulated images of the illumination conditions (incidence (i), emission (e), phase angle ($\alpha$)), and longitude (lon)/latitude (lat). For these images, a simple translation appears sufficient to ensure a good co-registration. Therefore, the phase cross-correlation algorithm available to the Python scikit-image package \citep{VanDerWalt_2014} was used to correct for the offset between original and synthetic images. 

\subsection{SRC data analysis} \label{sec:SRC_data_analysis}
\subsubsection{Observations at phase angle larger than 10°}
With the Super Resolution Channel (SRC), 3681 images of Deimos obtained from more than 300 observation sequences were acquired, with higher spatial resolution than HRSC. The typical resolution of the images ranges between 100 and 300 m/px. Some observations have a spatial resolution better than 100 m/px and up to 85 m/px. Only one observation has a resolution larger than 300 m/px. Time of observations covers more than 20 years, from October 2004 to December 2024; and the phase angle ranges from 0.06° to 138°. The images were visually inspected to verify their quality. We first remove from the dataset the images where Deimos is not fully in the field-of-view, and those that are saturated. In particular, the first image of an observation sequence is always used to adjust and optimize the exposure time and is very often saturated or too dark. After this first selection step, we obtained 2216 acceptable images. We then reduced the number of images with a selection of orbits covering the entire phase angle range with a sufficient phase angle step, and with the best possible spatial resolution. After this selection, we work on a dataset of 1245 Deimos images. The reduction of the number of data was important and necessary to decrease the computation time for the following reduction and analysis of the SRC data, in particular SPICE simulations and fitting.\par
As the SRC is not absolutely calibrated, the flux value for each pixel is provided in DN. In order to normalize the flux, it is necessary to divide by the exposure time [s] and multiply by the square of the heliocentric distance $r_h$ [AU] of the observation. The result is a normalized image provided in DN.s$^{-1}$. \par
We performed SPICE simulations and co-registration in a similar way to HRSC. A simple translation of the simulated images was sufficient to match the original images. \par
For disk-integrated analysis, the averaged flux was computed using aperture photometry, divided by the projected surface of Deimos. The aperture radius was determined individually and automatically for each image, based on the observation distance. The projected surface was estimated from the simulated images, therefore allowing to take into account for cast shadows. We then computed the average of the computed flux for a given orbit (therefore with almost the same phase angle) and the uncertainties of the flux comes from the standard deviation for the observation sequence. 

\subsubsection{The opposition effect dataset}
The opposition effect dataset obtained at small phase angles ($<$10°) exhibits peculiar characteristics compared to the other observation sequences. In contrast to the other observations, which were obtained in inertial observation mode and thus correspond to a limited number of images, the observations close to the opposition surge were obtained in spot tracking mode, where the pointing was maintained toward the center of Deimos while the spacecraft slewed for several seconds or minutes. Hence, the observation sequences are longer and contain a larger number of images than those obtained at larger phase angles. For example, the observation dataset during orbit K803 contains more than 80 images.\newline
As this type of observation is less common (only three for Deimos since the beginning of the mission), the data analysis was conducted in a slightly different manner. The three observation sequences at small phase angles were obtained during orbit L455, L568, and K803. K803 has the highest coverage in phase angle, ranging from 0.06° to 3.3°. Observation L455 covers between 0.5° and 0.9°, while L568 ranges from 0.8° to 2.1°. While L455 and L568 have similar flux level (1.35 $\times$ 10$^7$ ADU.\,$r_h^2$/s), K803 exhibits a lower flux (1.28 $\times$ 10$^7$ ADU.\,$r_h^2$/s). Even if this difference is small, it causes issues and leads to non-negligible uncertainties when trying to model the opposition surge. The discrepancy between the flux obtained during the three orbits may be related to the time of acquisition. L455 and L568 observations were obtained at a month of interval. K803 was obtained six months before the two other opposition observations. The flux variations are likely due to the modifications of the positions of Mars, MEX, and Deimos, and hence the presence of more or less straylight from Mars. In order to correct this issue, we computed the average of the flux at 1° and normalized the three opposition datasets to this value.\newline
The radiance factor values flatten at phase angles smaller than the angular size of the Sun as seen from Deimos \citep{Deau_2012, Fornasier_2024}. To avoid underestimating the opposition effect, we removed these data from our dataset.

\begin{figure*}
\resizebox{\hsize}{!}{\includegraphics{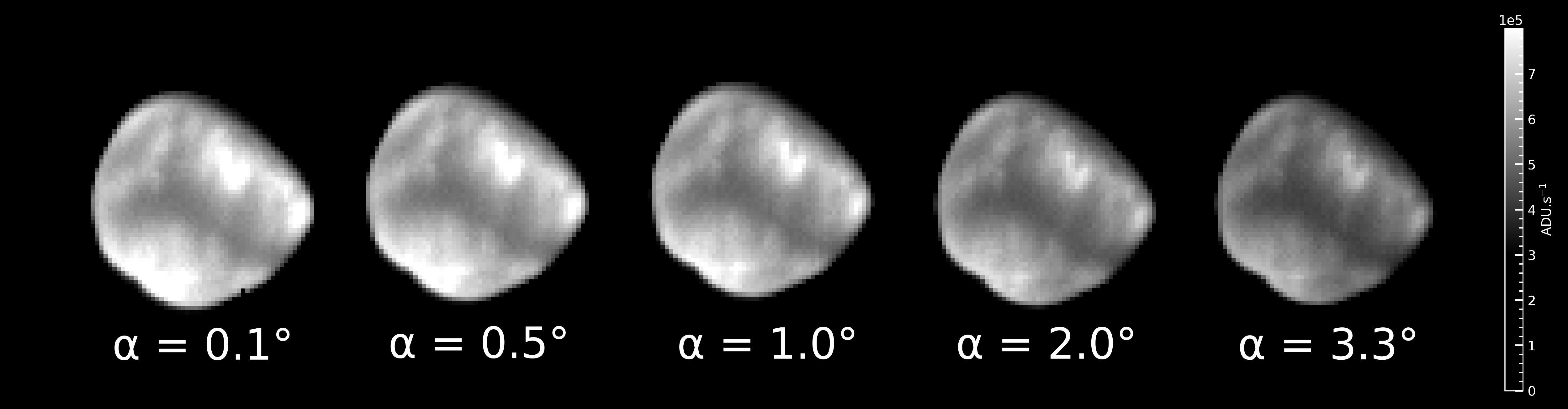}}
\caption{Deimos opposition effect observed by the SRC during orbit K803 on 2020-06-15. The flux is in ADU.s$^{-1}$; no other normalization was applied here (the heliocentric distance was the same for a given orbit).}
\label{fig:SRC_deimos_opposition_effect}
\end{figure*}

\section{Absolute calibration of the SRC}

\begin{table}[]
    \centering
    \caption{Parameters of the SRC \citep{Jaumann_2007, Oberst_2008}.}
    \begin{tabular}{cc}
    \hline
    \hline
    \textbf{SRC optical system} &\\
    \hline
    Focal length & 988.5 mm\\
    Diameter of the telescope aperture & 89.9 mm \\
    f-number & 11 \\
    \hline
    \textbf{SRC detector} &\\
    \hline
    Detector type & CCD\\
    Pixel size & 9 µm $\times$ 9 µm\\
    System gain & 5.3 e$^{-}$/DN\\
    \hline
    \end{tabular}
    \label{tab:SRC_properties}
\end{table}

\subsection{Mutual events observations}
The HRSC team has identified opportunities to observe Phobos and Deimos in conjunction with other celestial bodies. Specifically, SRC images of stars, Jupiter and the Galilean satellites, Saturn, Uranus, the Earth, and the Moon were captured. These images have so far been used for astrometric purposes, with the aim of improving ephemerides for the Martian moons \citep{Ziese_2018}.
However, certain observations are particularly useful for the absolute calibration of the SRC instrument, and our focus lies on the joint observations of the Martian Moons with stars and Jupiter. Jupiter is a well-known object that has been used to calibrate other cameras in the past, including the Halley Multicolor Camera onboard Giotto (HMC, \citealt{Thomas_1990}) and the Color and Stereo Surface Imaging System onboard TGO (CaSSIS, \citealt{Thomas_2022}).

\subsubsection{Jupiter}
Over a period of two decades (2004-2024), the SRC data have allowed us to identify several observations of Jupiter. Many observations of Jupiter are systematically saturated or close to saturation, in the non-linearity region of the detector. This makes them unsuitable for calibration. As a result, we have a set of two Deimos-Jupiter mutual events that are suitable. We present detailed characteristics of these observations, including distances, exposure time, and phase angle in Table \ref{tab:jupiter_observations_properties}.
Considering these observations of Deimos and Jupiter, we have a collection of 14 images of Jupiter that can be used to achieve absolute calibration. \newline 

\begin{table*}[]
    \centering
    \caption{Characteristics of mutual events Deimos-Jupiter. 
    }
    \resizebox{\textwidth}{!}{
    \begin{tabular}{cccccccccccc}
    \hline
    \hline
       Day & Start time\tablefootmark{a} & Stop time\tablefootmark{a} & MEX orbit n° & r$_h$\tablefootmark{b} & $\Delta$\tablefootmark{c} & $\delta$\tablefootmark{d} & $\alpha_J$\tablefootmark{e} & $\alpha_P$\tablefootmark{f} & $\tau$\tablefootmark{g} & No. images & No. used images\tablefootmark{h}\\
    \hline
       2018-01-28 & 20:42:55 & 20:43:22 & 17817 & 1.605 AU & 3.88 AU & 18 985 km & 5.65° & 32.31° & 20.160 ms & 8 & 7\\
       2018-02-01 & 15:33:43 & 15:34:10 & 17830 & 1.601 AU & 3.87 AU & 14 029 km & 5.04° & 62.41° & 20.160 ms & 8 & 7\\

    \hline
    \end{tabular}
    }
    \label{tab:jupiter_observations_properties}
    \tablefoot{
    Other mutual events Phobos-Jupiter were identified but are systematically saturated on Jupiter: 2018-08-30, 2018-09-06, 2018-09-30, 2020-01-25, 2020-08-10, and 2022-12-26. This is also observed for Deimos-Jupiter events: 2019-11-29, 2020-02-04, 2020-07-04, 2020-07-08, 2020-09-18, 2020-09-22, 2020-09-25, 2022-02-14.\\
    \tablefoottext{a}{Start and stop time correspond to the time of the beginning and end of the observation sequence.} \\
    \tablefoottext{b}{r$_h$ is the heliocentric distance of the spacecraft in AU.} \\
    \tablefoottext{c}{$\Delta$ is the MEX-Jupiter distance.} \\
    \tablefoottext{d}{$\delta$ is the MEX-Deimos distance.} \\
    \tablefoottext{e}{$\alpha_J$ is the phase angle in degrees between the Sun-MEX-Jupiter.} \\
    \tablefoottext{f}{$\alpha_P$ is the phase angle in degrees between the Sun-MEX-Phobos/Deimos.} \\
    \tablefoottext{g}{$\tau$ is the exposure time.} \\
    \tablefoottext{h}{Some images were discarded and not analyzed because of one of these reasons: (1) Phobos occults Jupiter, (2) the image is saturated.}
    }
\end{table*}

\subsubsection{Stars}
Observations were initially conducted for astrometry purposes and to check the pointing of the instrument. Based on pointing information in the header of each image, i.e., right ascension and declination, we were able to identify the stars present in the background of Phobos/Deimos images. These stars generally correspond to very bright stars with an apparent magnitude smaller than 5. Here, we consider recent observations of stars made by the SRC between 2020-02-01 and 2024-11-01. Unfortunately, none of the stars appears to be spectrophotometric standards. Some stars were saturated on one pixel and were removed from the potential list of the calibration targets. We also observed that some of the stars are variable stars or multiple star systems; therefore, not suitable for absolute calibration. The selected stars correspond to various spectral types from M-type to B-type, and various apparent magnitudes in the V filter from 5.5 to 1.6 (Table \ref{tab:star_observations_properties}).

\begin{table*}[]
    \centering
    \caption{Characteristics of mutual events Phobos/Deimos with stars.}
    \resizebox{\textwidth}{!}{
    \begin{tabular}{cccccccccc}
    \hline
    \hline
       Star & Day & Start time & Stop time & MEX orbit n° & mag$_{v}$ & Spectral type & $\tau$ & No. images & No. used images\tablefootmark{a}\\
    \hline
       $\beta$ Ophiuchi & 2020-02-11 & 11:16:17 & 11:17:26 & 20373 & 2.75 & K2III & 41.328 ms & 8 & 7\\
       $o$ Ursae Majoris & 2021-05-21 & 22:33:49 & 22:34:11 & 21972 & 3.42 & G5III & 32.336 ms & 8 & 7\\
       $\theta$ Ursae Majoris & 2021-08-20 & 20:43:02 & 20:43:31 & 22285 & 3.18 & F7V & 51.408 ms & 8 & 7\\
       $\kappa$ Ophiuchi & 2021-11-02 & 09:45:56 & 09:46:12 & 22538 & 3.20 & K2III & 50.400 ms & 8 & 7\\
       $\nu$ Ophiuchi & 2022-01-31 & 02:42:42 & 02:42:52 & 22846 & 3.34 & G9III & 27.720 ms & 8 & 7\\
       $\nu$ Ophiuchi & 2022-01-31 & 06:54:57 & 06:55:07 & 22846 & 3.34 & G9III & 32.256 ms & 8 & 6\\
       $\gamma$ Phoenicis & 2022-07-12 & 00:21:33 & 00:21:49 & 23402 & 3.42 & M0III & 45.360 ms & 8 & 7\\
       $\delta$ Andromeda & 2022-12-02 & 11:28:40 & 11:28:46 & 23895 & 3.28 & K3III & 49.896 ms & 8 & 4\\
       $\eta$ Aurigae & 2022-12-07 & 02:07:26 & 02:07:43 & 23911 & 3.18 & B3V & 24.192 ms & 8 & 7\\
       HD 36959 & 2024-06-13 & 06:45:09 & 06:45:25 & 25816 & 5.53 & B1V & 72.576 ms & 8 & 7\\
       HD 36960 & 2024-06-13 & 06:45:09 & 06:45:25 & 25816 & 4.72 & B0.5V & 72.576 ms & 8 & 7\\
       $\phi$ Eridani & 2024-07-12 & 10:33:29 & 10:33:52 & 25916 & 3.57 & B8IV & 54.936 ms & 8 & 7\\
       $\epsilon$ Gruis & 2024-07-19 & 10:52:48 & 10:53:05 & 25940 & 3.47 & A2IV & 61.992 ms & 8 & 7\\
       $\gamma$ Orionis & 2024-09-19 & 17:46:57 & 17:47:14 & 26154 & 1.64 & B2V & 32.256 ms & 8 & 7\\
       $\alpha$ Lyncius & 2024-09-28 & 17:21:58 & 17:22:14 & 26185 & 3.14 & K6III & 59.976 ms & 8 & 4\\
       $\delta$ Ursae Majoris & 2024-10-12 & 15:32:37 & 15:32:53 & 26233 & 3.32 & A2V & 68.544 ms & 8 & 6\\
       $\alpha$ Lyncius & 2024-11-01 & 19:23:02 & 19:23:16 & 26302 & 3.14 & K6III & 51.408 ms & 8 & 6\\
    \hline
    \end{tabular}
    }
    \label{tab:star_observations_properties}
    \tablefoot{
    \tablefoottext{a}{Some images were discarded and not analyzed because: (1) the star is too close to Phobos or Deimos, resulting in stray light contribution to the flux of the star, (2) the star is saturated on one or several pixels.}
    }
\end{table*}

\subsection{Absolute calibration procedure with mutual events observations}
\subsubsection{Data analysis: determination of the observed flux} \label{sec:flux_obs}
\begin{figure}
\centering
  \resizebox{\hsize}{!}{\includegraphics{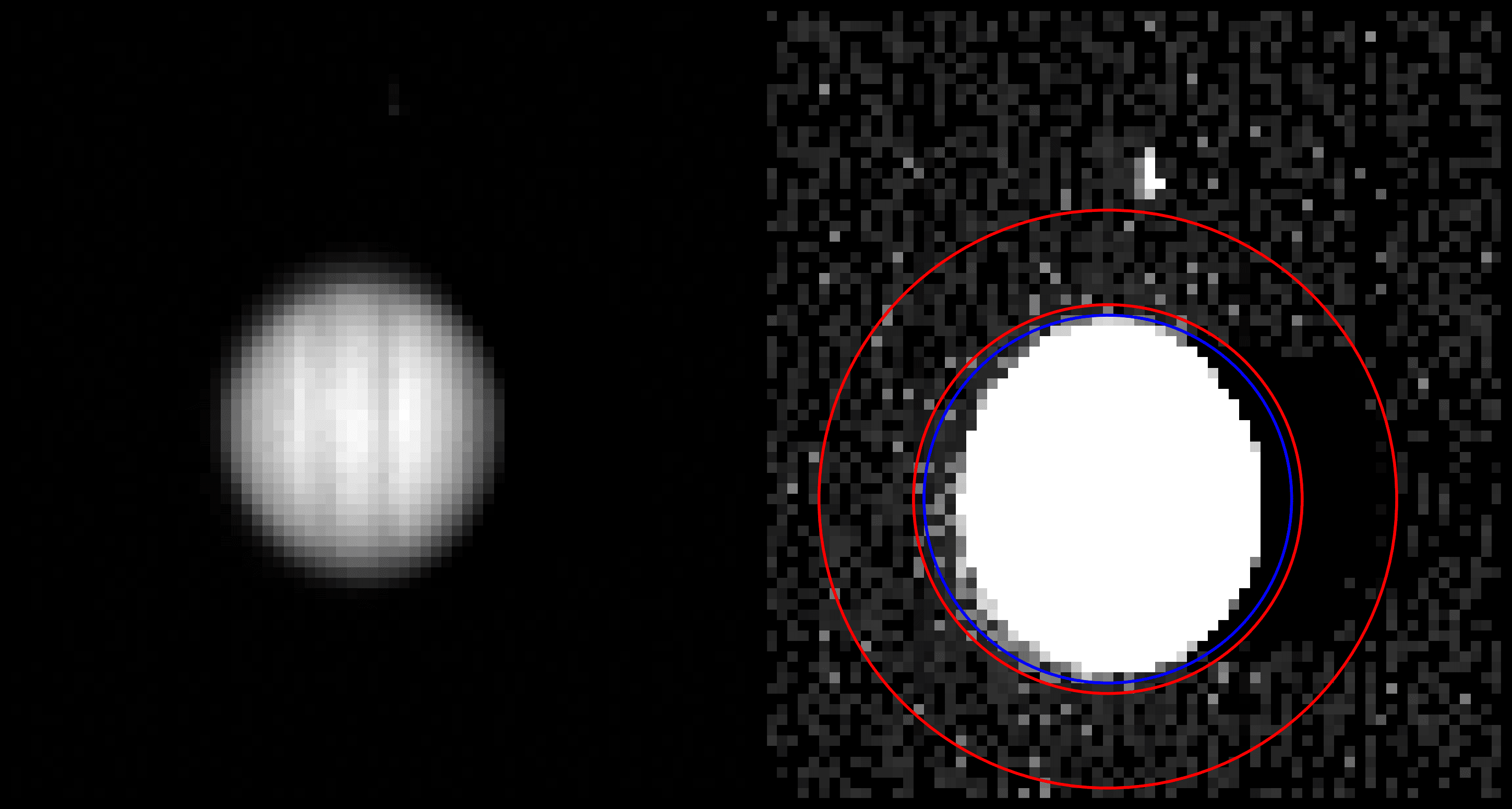}}
  \caption{An example of Jupiter SRC observation (2022-07-12T18:59:16) and the associated method to determine Jupiter's flux. Left: Image of Jupiter obtained by the SRC instrument. The bands of the atmosphere can be seen pretty well. Right: the same Jupiter image but clamped to 200 DN to show the background. The circle (blue) and the aperture annulus (red) are also shown. Note that the Io moon is present at the top of Jupiter. It is important to avoid the satellite for the aperture photometry method.}
  \label{jupiter_aperture_photo}
\end{figure}

For Jupiter, the aperture photometry technique is used to determine the total flux observed in digital number (Fig. \ref{jupiter_aperture_photo}). The total signal of Jupiter is calculated by integrating the measured signal within a circle of approximately 35 pixels in diameter (slightly larger than Jupiter's size with SRC in general). An annulus is subsequently defined, establishing the image background. The inner and outer radii of the annulus are selected to prevent the integration of Jupiter's flux and especially to avoid considering the flux of the moons that may appear in the images. Consequently, the annulus is defined as approximately five to ten pixels. Removing the background level is achieved by deducting it from the integrated flux while considering the number of pixels employed to integrate Jupiter's signal.
For stars, the aperture photometry was also used. However, as these bodies are not resolved and only visible as a PSF, the integration area was generally selected within a diameter of 8 pixels, and the background annulus between 15 and 20 pixels in diameter.

\subsubsection{Predicted flux}
The flux received from a source (considering a Lambert source) at a phase angle $\alpha$ is given by \citep{Tomasko_1976, kartunnen_1987}:
\begin{equation} \label{eq: solar_flux}
    F(\lambda) = F_{\odot}(\lambda) p(\lambda) \phi(\lambda, \alpha) \left( \frac{r^2 \times (1 \mathrm{AU})^2}{r_{h}^2 \Delta^2} \right),
\end{equation}
where $F_\odot$ is the solar flux received at 1 AU [W.\,m$^{-2}$.\,nm$^{-1}$], p is the geometric albedo, $\Phi$ is the phase function, r is the radius of the planet [AU], $r_h$ is the heliocentric distance [AU], and $\Delta$ is the observer-target distance [AU]. \par
The flux computed above is the ideal flux received by the observer. However, the instrumental characteristics (optics, detector, and filters responsivity) significantly diminish the incoming flux and should be considered. The responsivity of an imaging system can be modeled as \citep{Thomas_1990, Magrin_2015, Thomas_2022}:
\begin{equation}
    R = \frac{1}{G} \int_{0}^{\infty} F(\lambda) M(\lambda) Q(\lambda) T(\lambda) \frac{\pi d^{2}}{4} \frac{\lambda}{h c} \,d\lambda,
\end{equation}
where G is the detector gain [e$^{-}$/DN], d is the diameter of the aperture of the telescope [m], F is the input flux [W.\,m$^{-2}$.\,nm$^{-1}$], Q is the quantum efficiency of the detector [e$^{-}$/photon], and M and T are respectively the transmission of the telescope and of the filter. The responsivity is finally given here in [DN]. \par
We can therefore compute the ratio of the observed and predicted flux using the predicted responsivity computed with the equation above and using the flux determined from observation (Sect. \ref{sec:flux_obs}). \par
For stars, this last equation can be used directly, taking the input flux from the XP spectra of the Gaia DR3 catalog. For $\gamma$ Orionis, the spectrum was taken from \cite{Krisciunas_2017}. \par
In the case of Jupiter observations, we need to compute the theoretical flux of Jupiter. We followed the work of \cite{Thomas_2022} on the absolute calibration of the CaSSIS instrument. Similarly, we used the \cite{Meftah_2018} solar spectrum. The Jupiter's phase function was taken from \cite{Mayorga_2016}. Because phase functions were derived for different filters, we chose to use Jupiter's phase function obtained in the closest filter (in terms of central wavelength) to the SRC panchromatic filter. The RED filter central wavelength of the NAC ISS camera onboard CASSINI appears to be close to the SRC filter. For the geometric albedo spectrum, we used the full-disk albedo derived by \cite{Karkoschka_1998} with ground-based observations. The data were acquired at a phase angle of 6.8°. To obtain the geometric albedo spectrum, we corrected the \cite{Karkoschka_1998} data with the \cite{Mayorga_2016} phase function. All these data were interpolated at 1 nm within the range of the SRC panchromatic filter.

\subsubsection{Radiance calibration factor}
The spectral radiance $\phi$ [W.\,m$^{-2}$.\,nm$^{-1}$.\,sr$^{-1}$] can be computed using the formula from \cite{Magrin_2015} and is simply the average of Jupiter's radiance weighted by the SRC's instrumental response:
\begin{equation}
    \phi = \frac{\frac{1}{G} \int_{0}^{\infty} F(\lambda) M(\lambda) Q(\lambda) T(\lambda) \frac{\pi d^{2}}{4} \frac{\lambda}{h c} \,d\lambda}{\Omega \frac{1}{G} \int_{0}^{\infty} M(\lambda) Q(\lambda) T(\lambda) \frac{\pi d^{2}}{4} \frac{\lambda}{h c} \,d\lambda},
\end{equation}
where $\Omega$ is the pixel size in steradians, i.e. $\Omega$ = $\left(\frac{p}{f}\right)^{2}$, with $p$ corresponding to the pixel size and $f$ to the focal length. For the SRC, $\Omega = 8.289 \times 10^{-11}$ sr. This equation can be simplified because $G$, $d$, and $hc$ are not wavelength-dependent:
\begin{equation} \label{eq: spec_radiance}
    \phi = \frac{\int_{0}^{\infty} F(\lambda) M(\lambda) Q(\lambda) T(\lambda) \lambda \,d\lambda}{\Omega \int_{0}^{\infty} M(\lambda) Q(\lambda) T(\lambda) \lambda \,d\lambda}
\end{equation}
Then, the absolute calibration factor to obtain radiance from the count rate image is simply given by the ratio of the observed total count rate divided by the spectral radiance: 
\begin{equation}
    A_{radiance} = \frac{R}{\phi}
\end{equation}
This equation can be developed using Eqs. (\ref{eq: solar_flux}) and (\ref{eq: spec_radiance}):
\begin{equation} \label{eq:calib_abs}
    A_{radiance} = \frac{R \Omega \int_{0}^{\infty} M(\lambda) Q(\lambda) T(\lambda) \lambda \,d\lambda}{\int_{0}^{\infty} F(\lambda) M(\lambda) Q(\lambda) T(\lambda) \lambda \,d\lambda}
\end{equation}
The obtained absolute calibration factor $A_{radiance}$ is given in DN/s.(W.\,m$^{-2}$.\,nm$^{-1}$.\,sr$^{-1}$)$^{-1}$.

\subsubsection{I/F calibration factor} 
The expected solar count rate can be computed using the following equation: 
\begin{equation}
    R_\odot = \frac{1}{G} \int_{0}^{\infty} F_\odot(\lambda) M(\lambda) Q(\lambda) T(\lambda) \frac{\pi d^{2}}{4} \frac{\lambda}{h c} \,d\lambda,    
\end{equation}
and the solar spectral irradiance at 1 AU is expressed by:
\begin{equation}
    \phi_\odot = \frac{\int_{0}^{\infty} F_\odot(\lambda) M(\lambda) Q(\lambda) T(\lambda) \lambda \,d\lambda}{\int_{0}^{\infty} M(\lambda) Q(\lambda) T(\lambda) \lambda \,d\lambda},
\end{equation}
The radiance calibration factor $A_{radiance}$ can then be used for the determination of the I/F calibration factor $A_{I/F}$:
\begin{align} 
    A_{I/F} &= A_{radiance} \times \frac{\phi_\odot}{\pi} \\
            &= \frac{R \Omega}{\pi}\frac{\int_{0}^{\infty}  F_\odot(\lambda) M(\lambda) Q(\lambda) T(\lambda) \lambda \,d\lambda}{\int_{0}^{\infty} F(\lambda) M(\lambda) Q(\lambda) T(\lambda) \lambda \,d\lambda} \label{eq:calib_IF}
\end{align}

\subsubsection{Color correction factor}
As we considered observations of two different bodies, obtained with different instruments, it is also essential to include a color correction factor that takes into account for the differences in albedo spectra. For Jupiter, the factors were computed by using the following: 
\begin{equation} \label{eq:color_corr_1}
    C = \frac{\int_{0}^{\infty}  p_{obs}(\lambda) M(\lambda) Q(\lambda) T(\lambda) \,d\lambda}{\int_{0}^{\infty} p_{cal}(\lambda) M(\lambda) Q(\lambda) T(\lambda) \,d\lambda} \times \frac{p_{cal, 600 nm}}{p_{
    obs, 600 nm}}
\end{equation}
where $p_{obs}$ and $p_{cal}$ are the albedos of, respectively, the main target of the observation (i.e., Deimos in this paper) and Jupiter.
For stars, we took into account the variability of the Sun spectra with stars with different spectral types:
\begin{equation} \label{eq:color_corr_2}
    C = \frac{\int_{0}^{\infty}  F_{\odot}(\lambda) M(\lambda) Q(\lambda) T(\lambda) \,d\lambda}{\int_{0}^{\infty} F_{*}(\lambda) M(\lambda) Q(\lambda) T(\lambda) \,d\lambda} \times \frac{F_{*, 600 nm}}{F_{
    \odot, 600 nm}}
\end{equation}

\subsubsection{Definition of synthetic filters}
The SRC is equipped with a panchromatic filter with a central wavelength of 600 nm and a bandwidth of 250 nm. However, due to time constrain before the launch of Mars Express, no full ground-based calibration was made. Therefore, the transmission of the filters and the optics of the telescope remains unknown. This information is needed for the absolute calibration process. \par
\begin{figure}
\resizebox{\hsize}{!}{\includegraphics{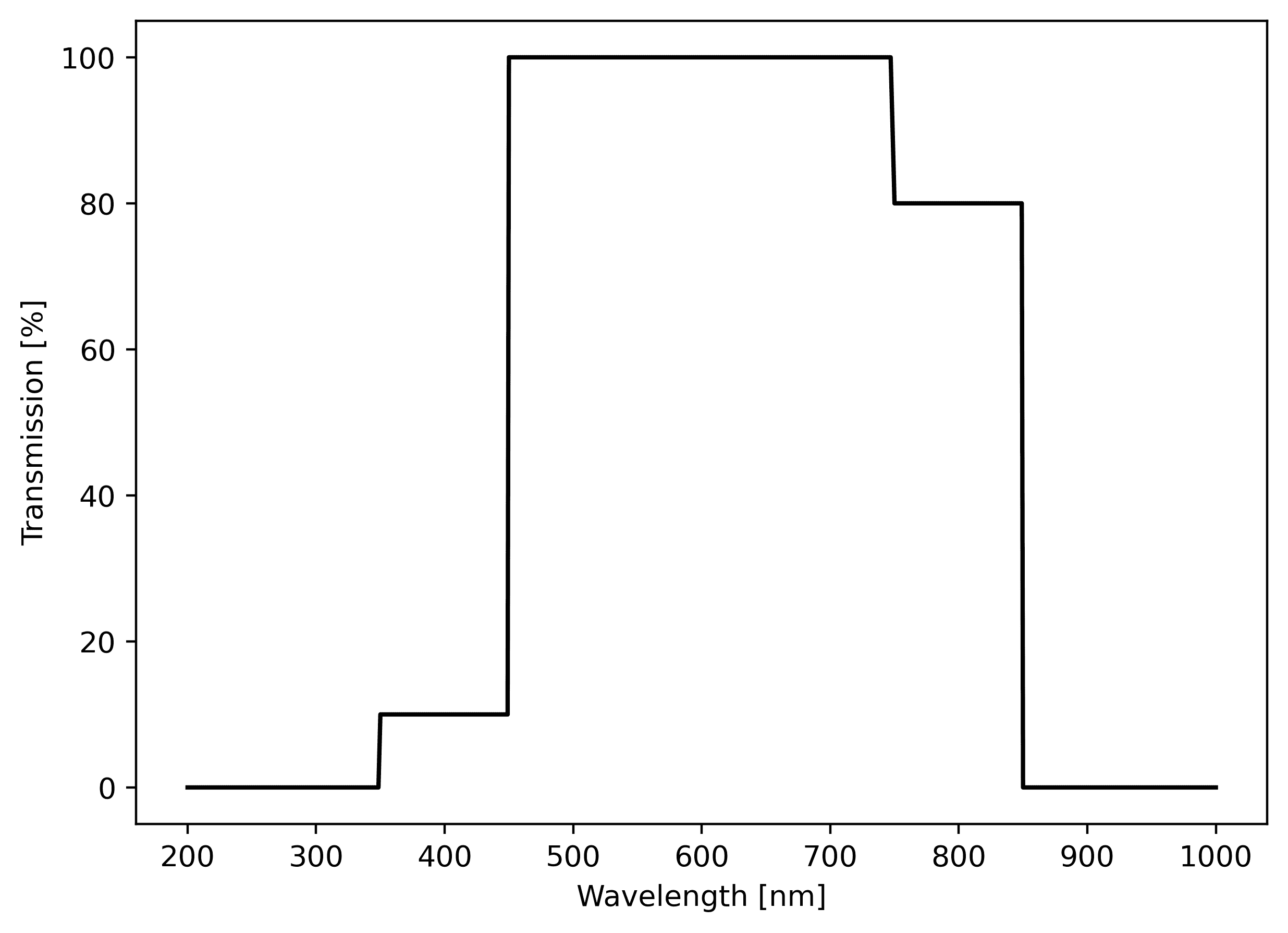}}
\caption{Synthetic filter used for the absolute calibration of the SRC.}
\label{fig:synthetic_filter}
\end{figure}

We defined synthetic filters covering the known central wavelength and bandwidth to perform the absolute calibration. As we have no information about the transmissions, we first tried to run our SRC calibration routine with wavelength-independent transmissions (i.e., constant on the bandwidth) at different levels, from a filter transmission of 1 to 0.5 and with optics transmission from 0.7 to 0.5. However, when looking at the calibration of stars, we found an important correlation between the absolute calibration factor and the temperature of stars: the coldest stars tend to have a higher calibration factor. This observation told us that the flux at short wavelengths is overestimated for B- and A-types because the maximum flux for these stellar types peaks in the UV or at the beginning of the visible. We then tried to minimize the flux coming from the short wavelength while increasing the flux coming from the longer wavelength by setting a wavelength-dependent filter. The filter is shown in Fig. \ref{fig:synthetic_filter}. As we have no information on the filter, we kept the synthetic filter simple. This filter is also more reliable than one with constant transmission, and closer to existing panchromatic filters. With this filter, the correlation between stars' temperature and calibration factor derived for each individual star is removed, and we observed a similar calibration factor for the different stellar types (within uncertainties). The total integrated transmission of the filter is 0.78. We used a constant telescope transmission of 0.7, leading to a total transmission of the optical parts of 0.55, consistent with the estimation of 0.5 given by \cite{Oberst_2008}. \par
It is also important to note that the transmissions do not substantially modify the final calibration factors. For example, we tried to run the calibration routine with two different constant total transmissions of 0.36 and 0.7 and found an I/F calibration factor of 1.85 $\times$ 10$^{7}$ and 1.82 $\times$ 10$^{7}$ DN/s, respectively. Of course, we also tried using a non-constant filter transmission, and the results were similar, with no significant changes. 

\subsubsection{Observed vs. Expected signals}
\begin{figure}
\resizebox{\hsize}{!}{\includegraphics{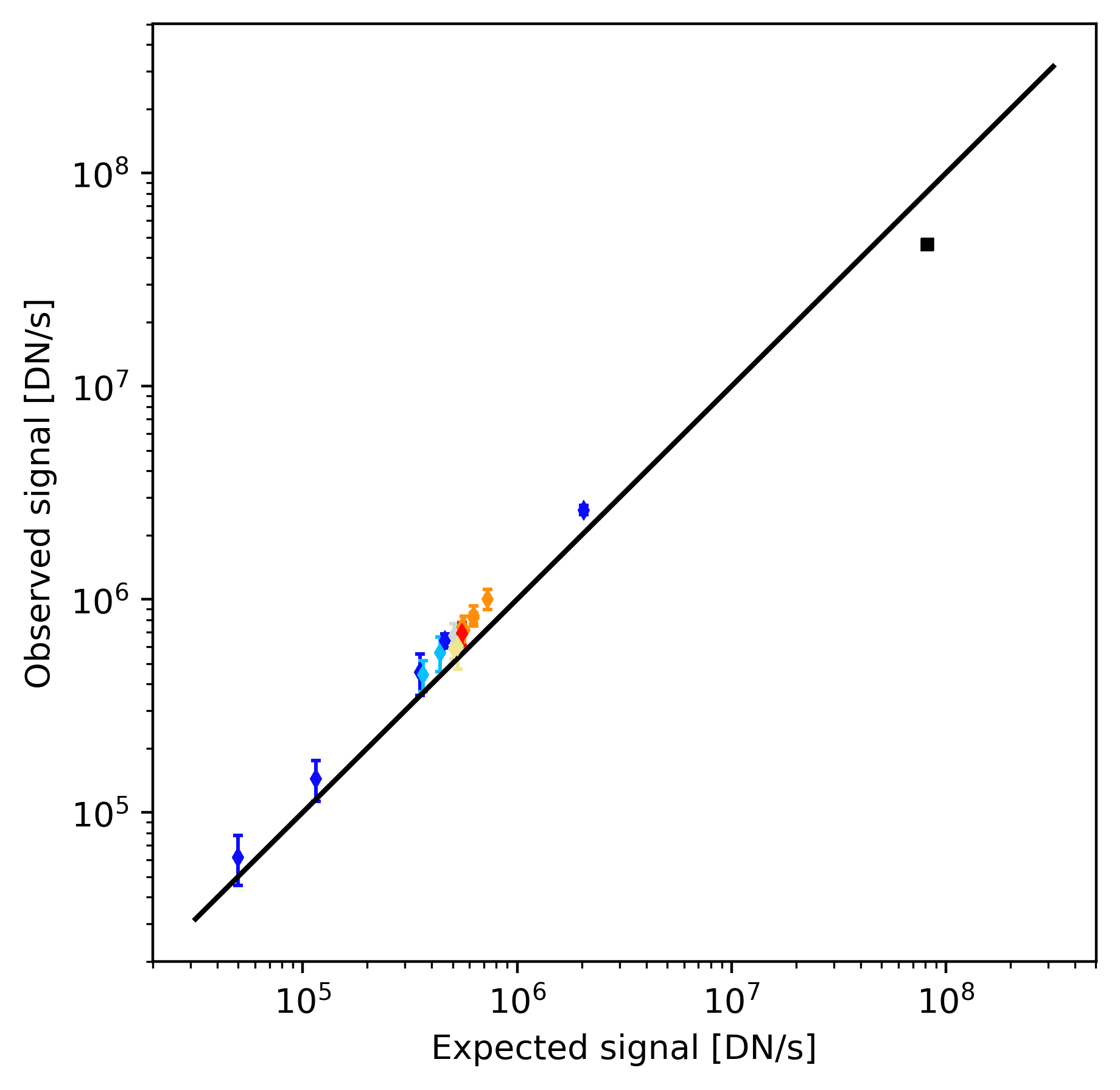}}
\caption{Observed vs. expected signal for each object/method. The black squares represent Jupiter data. The diamond shapes correspond to the stars used in this study, with the different colors representing the different spectral types. The black solid line represents an ideal camera with a ratio observed/expected equal to 1.}
\label{fig:expected_vs_observed}
\end{figure}

After finding a coherent and possible synthetic filter for the SRC, we can compute both observed and expected signals for the stars and Jupiter (Fig. \ref{fig:expected_vs_observed}). The observed flux is derived directly from the images by performing aperture photometry, and the expected flux is obtained for stars from the Gaia DR3 data, and for Jupiter from the Equation \ref{eq: solar_flux}. We can observe that the flux of stars is slightly higher than the expected flux, while the Jupiter observations exhibit a smaller flux compared to the expected flux. The uncertainties associated with each star or with each Jupiter observation correspond to the variations in flux coming from the images of the observation sequences (generally 7 images).\par
Furthermore, a potential correlation between the time of observation and the deviation of the ratio of observed flux to expected flux was investigated. This is a crucial aspect to consider if there has been a significant decrease in the performance of the SRC over the past two decades since its launch, specifically with regard to the quantum efficiency of the detector, which is likely to decline over time. However, no evidence was found to support a link between time and the performance of the SRC. Consequently, a single absolute calibration factor is adopted for the entire observation period.

\subsection{Absolute calibration procedure with HRSC calibrated filters data}
In order to compare with the mutual events calibration procedure, we also considered a calibration method using the absolutely calibrated HRSC color filters, which individually cover partially the range of the SRC panchromatic filter. \newline
The I/F calibration factor is computed from a linear regression of the ratio between the SRC signal [ADU.$r_h^2$/s] and the HRSC signal [I/F] (both computed from aperture photometry) and the phase angle. The derived SRC and HRSC signals correspond to observations of Deimos taken with a similar phase angle.\newline
If this method presents the advantage to compare directly with calibrated data of the same bodies taken in relatively similar conditions and do not require the transmission of the filter, they also have the disadvantage to not represent the same spectral behavior, with the four HRSC filters that do not represent the same spectral range and the same transmission as the SRC panchromatic filter. \par
For each HRSC filter, we computed the disk-integrated radiance factor and the associated disk-integrated SRC flux [DN.\,s$^{-1}$]. From the HRSC and SRC fluxes, we computed a linear fit to derive the calibration factors in the different filters (Fig. \ref{fig:HRSC_vs_SRC}). It is important to note that the HRSC calibration factor issue of the red and IR filters does not affect our results, because it results only in a shift of the HRSC values but does not modify the slope (i.e., the SRC calibration factor).

\subsection{Results of the absolute calibration}
From Equations \ref{eq:calib_abs}, \ref{eq:calib_IF}, \ref{eq:color_corr_1}, and \ref{eq:color_corr_2}, the absolute calibration factors can be computed. The color correction factors $C$ were employed to compute the I/F calibration factor. These factors are then multiplied by $A_{I/F}$ to produce the final I/F calibration factor. The final calibration factors are computed from the average of the calibration factors derived for each method/object presented in Table \ref{tab:absolute_cal_results}. Finally, the I/F correction factor used in the following is:
\begin{equation}
    A_{I/F} = (1.73 \pm 0.13) \times 10^{7} \quad \textrm{DN/s}
\end{equation}
We estimated an error in the absolute calibration of the SRC of about 7.5\% (1$\sigma$) based on the uncertainties of a single object (e.g., a star) during a sequence of observations and the uncertainties between different objects/methods. This last source of uncertainty is also related to the uncertainties associated with the transmissions of the filter and the other optical components. \newline
The factor derived in this work is not exactly the same as the reflectance scaling factor given in the header of HRSC image files. The factor, here, has to be understood as follows: (1) an SRC image must be divided by their exposure times, (2) multiplied by the square of the heliocentric distance in AU, (3) and divided by the absolute calibration factor to convert from count rate (DN/s) into I/F for the SRC.

\section{Disk-integrated photometry}
We performed disk-integrated photometry in a similar way to \cite{Fornasier_2024}: we used the disk-integrated Hapke model \citep{Hapke_1993} with a single-term Henyey-Greenstein (1T-HG) phase function to model the integrated flux obtained at different illumination angles. We additionally tried to fit the disk-integrated Deimos flux with a two-term Henyey-Greenstein (2T-HG). As the surface of Deimos is dark and to avoid too many free parameters for the disk-integrated analysis, we neglect the coherent backscattering opposition effect (CBOE, \citealt{Shevchenko_2012}). Therefore, we considered five free parameters in the model: the single-scattering albedo (SSA) $\omega_{\lambda}$, the asymmetry parameter $g_{\lambda}$, the shadow-hiding opposition effect (SHOE) parameters $B_{sh,0}$ and $h_{sh}$ -- representing respectively the intensity and the half-width of the opposition surge --, and the average roughness angle $\bar{\theta}$. The equations used are presented in Appendix \ref{appendix:hapke}. 

\subsection{HRSC}

\begin{figure*}
\centering
  \resizebox{\hsize}{!}{\includegraphics{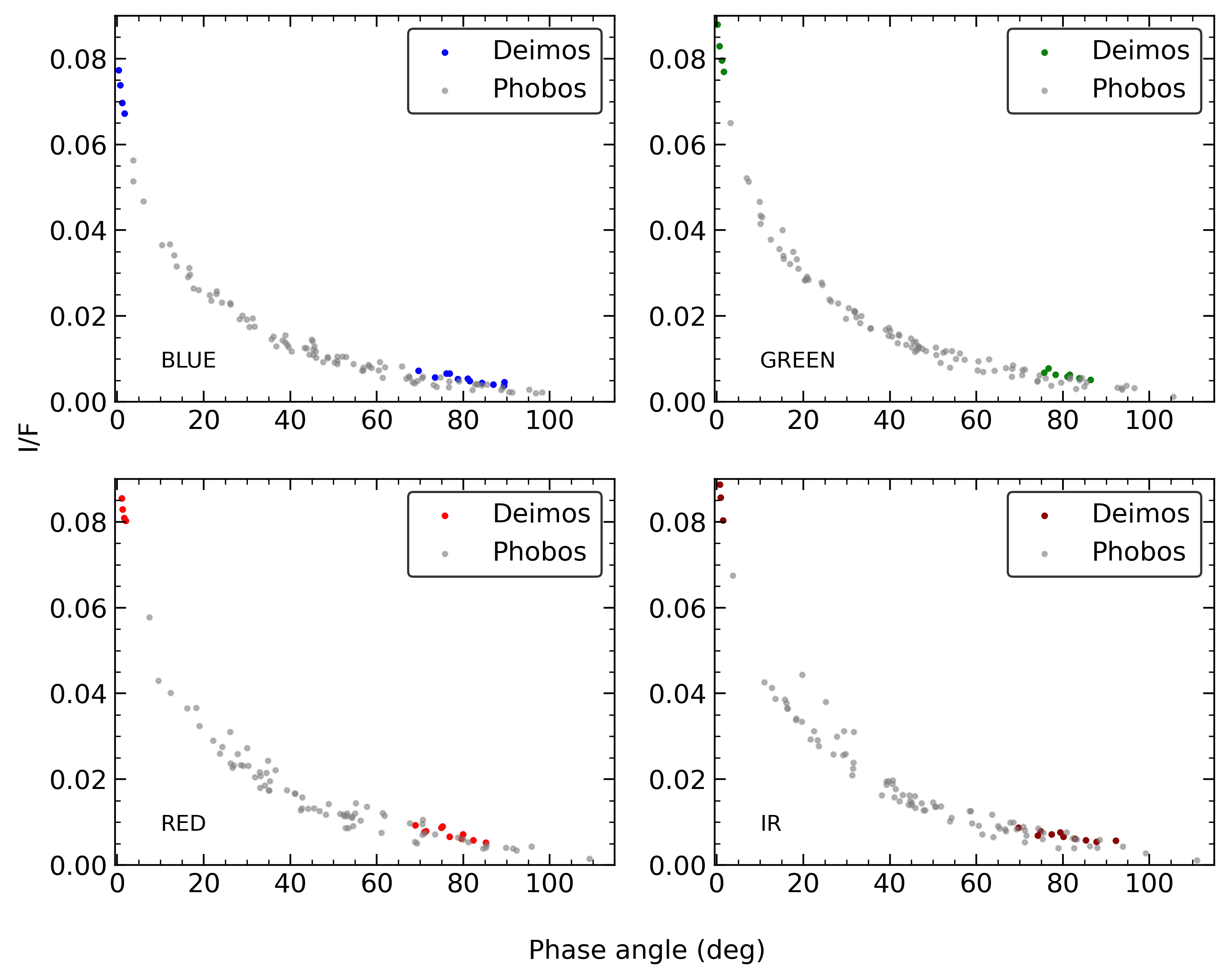}}
  \caption{Disk-integrated Deimos phase curve in the four HRSC filters. Grey points are the Phobos phase curve for comparison, derived from the HRSC camera in \cite{Fornasier_2024}.}
  \label{fig:disk-integrated_HRSC}
\end{figure*}

HRSC photometry of Deimos appears to be quite limited because of the low resolution of the images and because of the small range of phase angles for the images. Disk-integrated photometry was performed using the aperture photometry method to compute the integrated flux and using the projected surface derived from SPICE simulations. The Deimos phase curve is limited to small phase angles and high phase angles only, with a large lack of data between 5° and 70°. With the HRSC data, Deimos has similar phase curves in the four filters to Phobos. The similarity is particularly important in the red and IR filters, where the observations appear to overlap the Phobos' observations taken at the same phase angle. In the blue and green filters, Deimos is slightly brighter than Phobos (Fig. \ref{fig:disk-integrated_HRSC}). We were not able to perform Hapke modeling here because of the lack of phase angle coverage.

\subsection{SRC}
\begin{figure*}
    \centering
     \begin{subfigure}[b]{0.49\textwidth}
         \centering
         \resizebox{\hsize}{!}{\includegraphics[width=\textwidth]{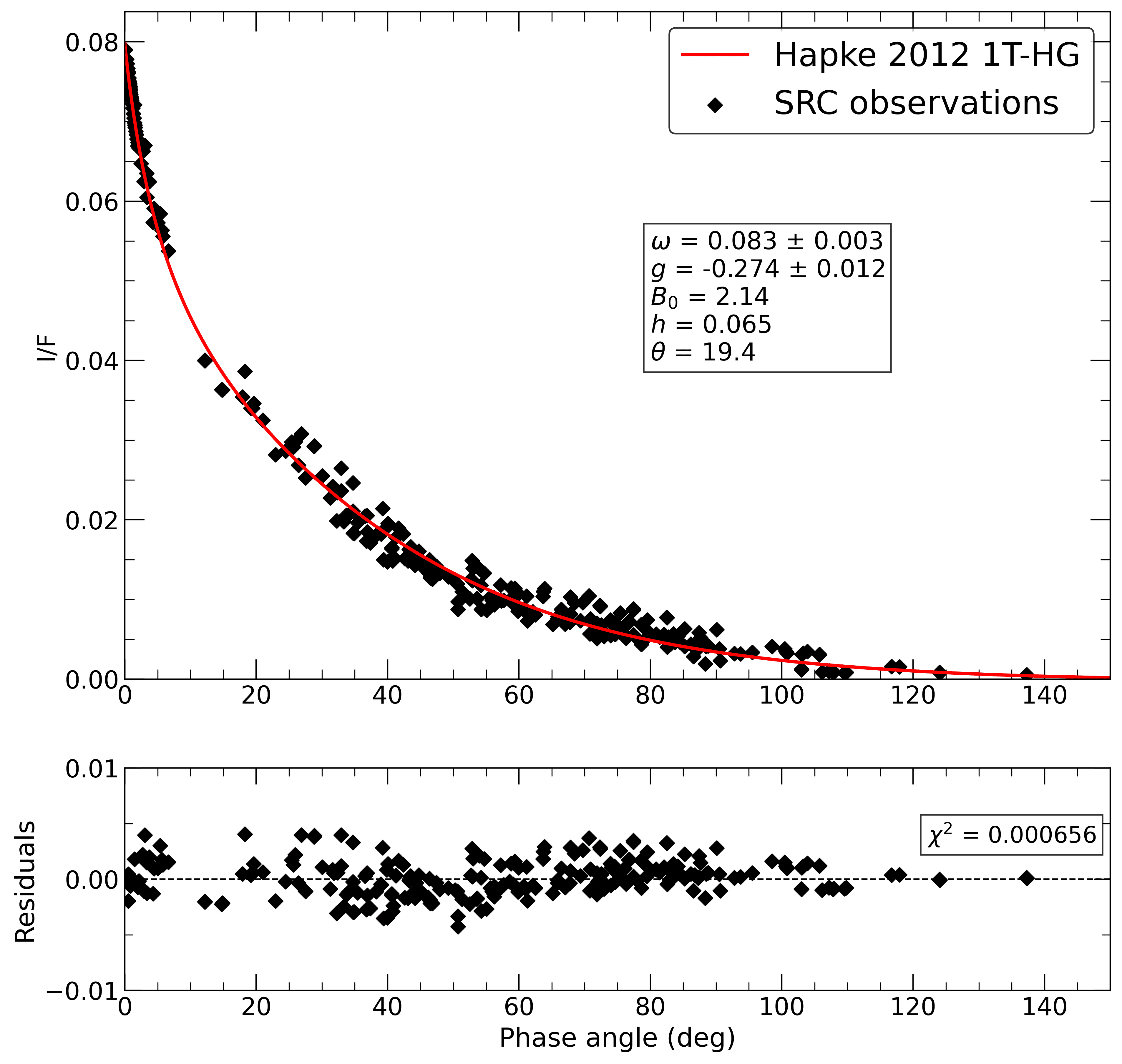}}
         \caption{}
         \label{fig:SRC_deimos_disk-integrated_phase_curve1}
     \end{subfigure}
     \hfill
     \begin{subfigure}[b]{0.49\textwidth}
         \centering
         \resizebox{\hsize}{!}{\includegraphics[width=\textwidth]{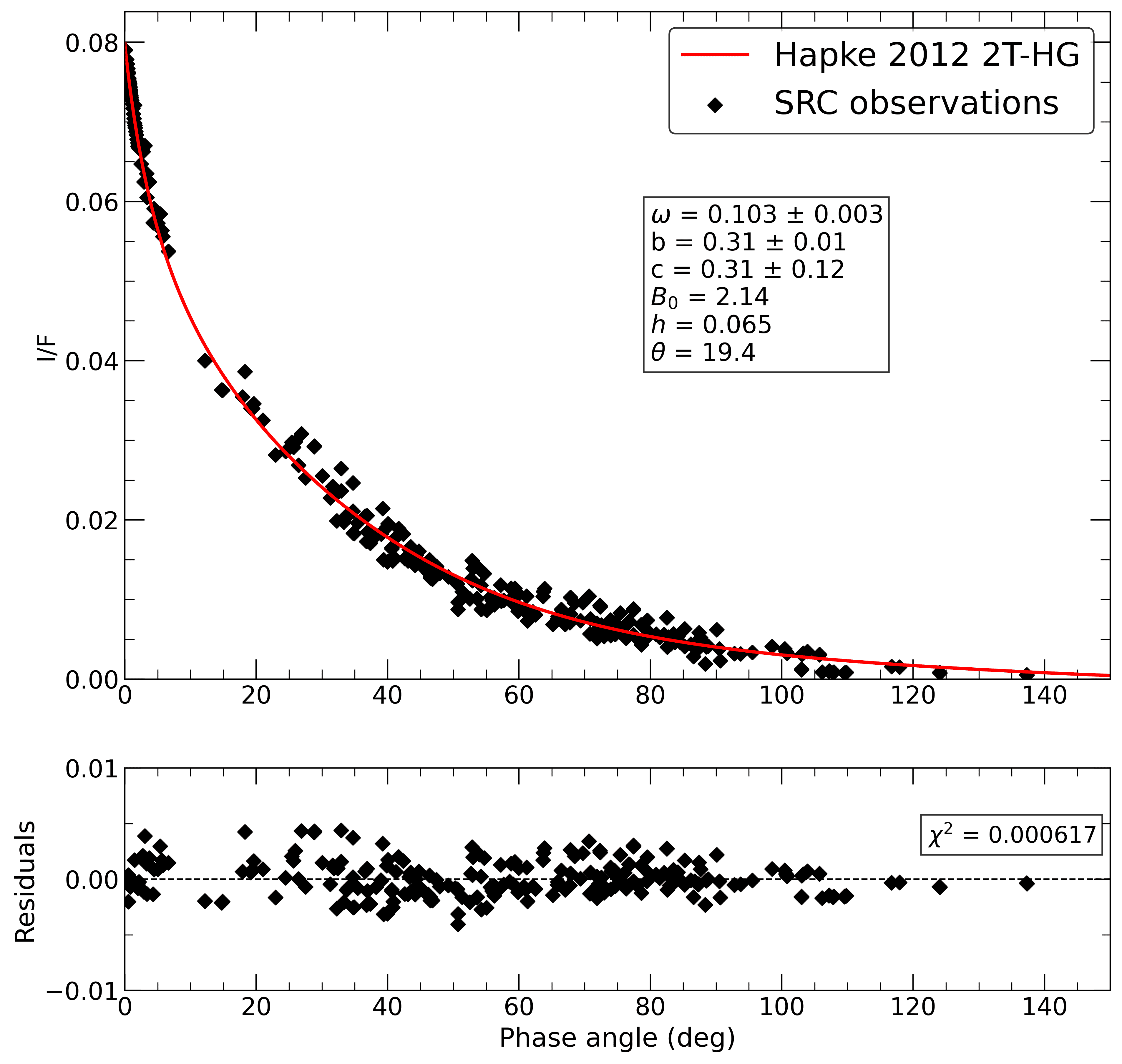}}
         \caption{}
         \label{fig:SRC_deimos_disk-integrated_phase_curve2}
     \end{subfigure}
     \caption{SRC disk-integrated phase curve (black diamond points) with (a) the Hapke 1T-HG global fit and (b) the Hapke 2T-HG global fit (red solid line). The residuals of the fit are given in the bottom subplot.}
     \label{fig:SRC_deimos_disk-integrated_phase_curve}
\end{figure*}

SRC observations are particularly interesting because they have a large coverage of phase angle and because they cover the opposition effect (0.06-120°). After performing aperture photometry, we obtain the following phase curve (Fig. \ref{fig:SRC_deimos_disk-integrated_phase_curve}). Because of the few scattered data available at large phase angles, we need to fix the $\bar{\theta}$ parameter to converge to a reliable solution with reasonable uncertainties. The value was determined afterward when performing disk-resolved Hapke modeling (see Sect. \ref{sec:disk_resolved}) and therefore set to $\bar{\theta} = 19.4$° for 1T-HG and $\bar{\theta} = 21.1$° for 2T-HG. The opposition effect parameters were also fixed to the values derived from disk-resolved Hapke modeling. We used the Levenberg-Marquardt algorithm to fit the Hapke model to our observations. The initial parameters of the free parameters were set to the values from \cite{Thomas_1996}. The results are presented in Table \ref{tab:comp_disk_integrated_analysis_model}. \par
We observed that the use of a double-term Henyey-Greenstein phase function leads to a better fit of the disk-integrated data. However, the obtained single-scattering albedo is much higher than expected, while the SSA derived from the single-term Henyey-Greenstein phase function is more reliable ($\omega = 0.083 \pm 0.003$). This value is consistent with the SSA derived by \cite{Thomas_1996} $\left(\omega = 0.079_{-0.006}^{+0.008}\right)$. The asymmetry parameter ($g = -0.274 \pm 0.012$) is also similar within uncertainties to the one in \cite{Thomas_1996}.

\begin{table*}[]
    \centering
    \caption{Hapke parameters found from the SRC disk-integrated analysis using the Hapke model with the single-term Henyey-Greenstein function and with the double-term Henyey-Greenstein function.}
    \resizebox{\textwidth}{!}{
    \begin{tabular}{ccccccccccc}
    \hline
    \hline
       Model & $\omega$ & $g$ or $b$ & $c$ & $B_{sh,0}^\dagger$ & $h_{sh}^\dagger$ & $\bar{\theta}^\dagger$ [deg] & $A_p$ & $A_B$ & $q$ & RMS\\
    \hline
       H2012-1THG\tablefootmark{a} & 0.083 $\pm$ 0.003 & -0.274 $\pm$ 0.012 & -- & 2.14 & 0.065 & 19.4 & 0.080 $\pm$ 0.001 & 0.018 $\pm$ 0.001 & 0.228 $\pm$ 0.002 & 0.00148\\
       H2012-2THG\tablefootmark{b} & 0.103 $\pm$ 0.003 & 0.31 $\pm$ 0.02 & 0.31 $\pm$ 0.12 & 2.14 & 0.065 & 21.2 & -- & 0.023 $\pm$ 0.001 & -- & 0.00144\\
    \hline
    \end{tabular}
    }
    \label{tab:comp_disk_integrated_analysis_model}
    \tablefoot{\\
    $\dagger$ $B_{sh,0}$, $h_{sh}$, and $\bar{\theta}$ were fixed based on the results of the SRC disk-resolved analysis with either 1T-HG or 2T-HG results.\\
    \tablefoottext{a}{H2012-1THG: Hapke 2012 model with porosity factor, single-term Henyey Greenstein phase function}\\
    \tablefoottext{b}{H2012-2THG: Hapke 2012 model with porosity factor, double-term Henyey Greenstein phase function}\\
    }
\end{table*}

\section{Disk-resolved spectro-photometry} \label{sec:disk_resolved}
\subsection{Color surface reflectance variations with HRSC}
\begin{figure}
\resizebox{\hsize}{!}{\includegraphics{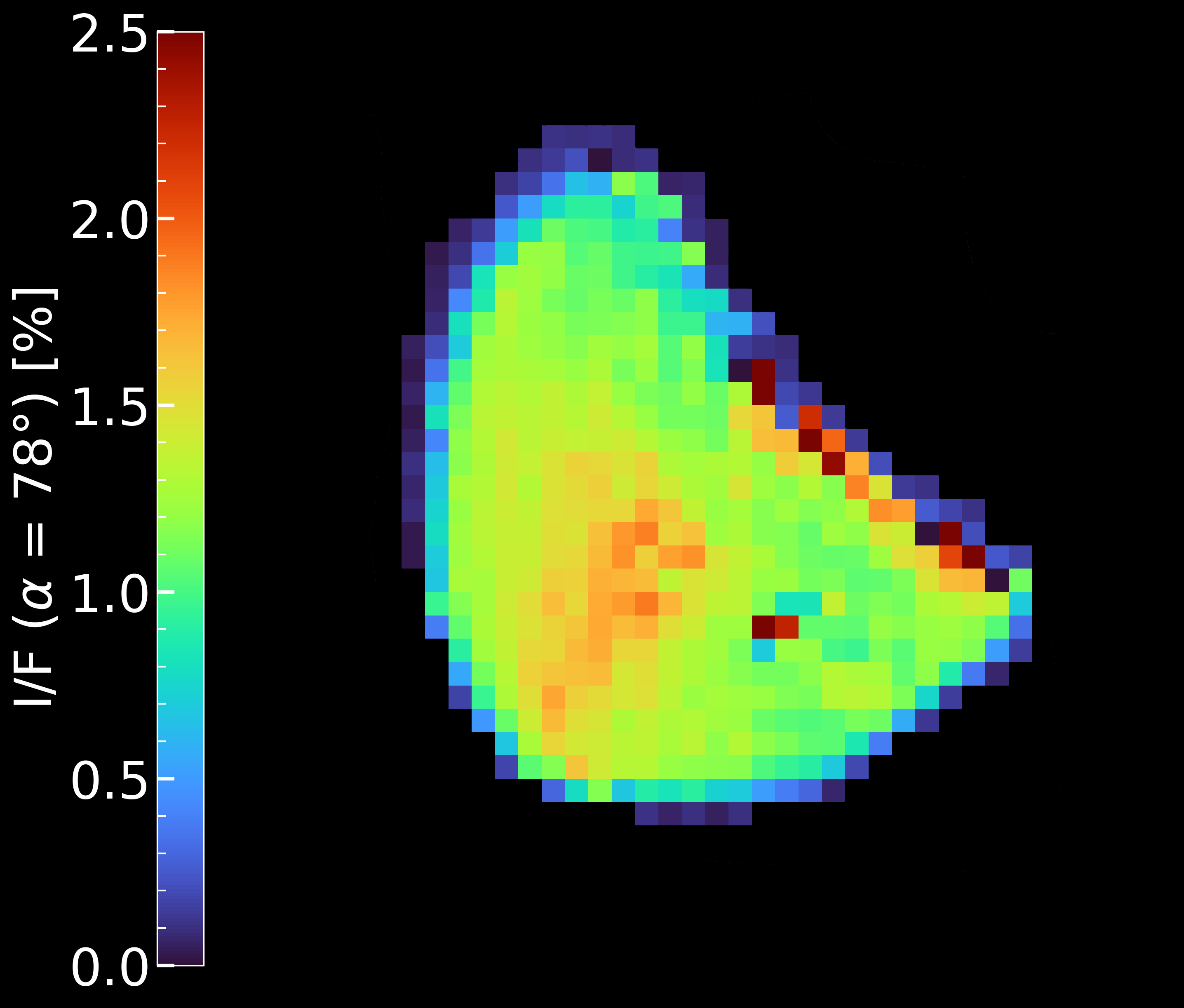}}
\caption{Example of I/F images of Deimos for one HRSC observation in the green filter. The I/F value is corrected for the illumination conditions using the Lommel-Seeliger law. The edges of Phobos are not physical and are residuals of the disk-function correction.}
\label{fig:HRSC_deimos_disk-resolved}
\end{figure}

Although the resolution of the images taken by the HRSC is low, disk-resolved analysis can also be performed. For this purpose, we chose the image with the best spatial resolution and with the smallest phase angle in the green filter (2021-04-28). The image is corrected for illumination conditions using the Lommel-Seeliger disk-function (Fig. \ref{fig:HRSC_deimos_disk-resolved}). We can notice that one region in the center appears brighter. It corresponds to a ridge clearly visible in other Deimos images. Two pixels are also particularly bright near the ridge. The position suggests that these pixels correspond to an impact crater.

\subsection{Spectro-photometric properties of the surface of Deimos}
The disk-resolved photometry with the SRC data was performed using the Hapke IMSA model (\citealt{Hapke_2012}, see Appendix \ref{appendix:hapke}), as well as the Kaasalainen-Shkuratov (KS) model \citep{Kaasalainen_2001, Shkuratov_2011}. For the disk-resolved analysis, we considered the use of several versions of the Hapke model with single- or double-term Henyey-Greenstein phase function, and with the addition of the coherent-backscattering opposition effect (CBOE). We also considered three different KS models with different disk functions for each (Lunar-Lambert (McEwen), Minnaert, and Akimov, see Appendix \ref{appendix:KS}). After co-registration of the simulated images (Sect. \ref{sec:SRC_data_analysis}), we created cubes of data containing the original image, incidence, emission, phase, latitude, and longitude images. For each data cube, a filtering procedure was applied to the pixels, with specific constraints on the illumination angles. This process was undertaken to avoid extreme and unfavorable observations. In the case of images exhibiting a mean phase angle greater than 5 degrees, the Hapke or KS modelling is applied exclusively to pixels characterised by incidence and emission angles smaller than 70° and a phase angle greater than 10\% of the maximum phase angle observed. For images with a mean phase angle that is less than 5°, the condition is that the phase angle should be greater than 0.01 degrees, with incidence and emission no greater than 70°. Following the exclusion of these pixels, the disk-resolved modelling on Deimos was performed on a dataset containing more than three million pixels. We decided to bin these data in order to (i) make the Hapke inversion procedure more efficient, and (ii) avoid having too much weight in the phase angle range where many observations were made. The binning procedure entailed the division of the phase angle (10°-140°) into 65 distinct bins, with each bin representing two degrees of phase angle. For each phase angle bin, a random sample of approximately 3000 data points is extracted. In case the number of points in a bin is less than the defined sample size, all points are retained. We have kept all the data at the opposition (0-10°). The minimization and fit of the Hapke IMSA model was performed using a Levenberg-Marquardt algorithm as for the disk-integrated analysis. The minimization procedure can not be run with all free parameters, because the Hapke model will well fit the data with the five (or up to eight) free parameters, but will give an unreliable solution as many parameters are correlated with each other. Therefore, we applied the following procedure when fitting the Hapke model: 
\begin{enumerate}
    \item We first inversed the Hapke model with the entire dataset, with all parameters free, and searching solutions within the following boundaries: $\omega$ = \{0.01,0.3\}, $g$ = \{-1.0,1.0\}, $B_{sh,0}$ = \{0.0,3.0\}, $h_{sh}$ = \{0.0,0.15\}, $\bar{\theta}$ = \{5°,50°\}, and depending on the case $B_{cb,0}$ = \{0.0,1.0\}, $h_{cb}$ = \{0.0,1.0\}. The values derived from disk-integrated photometry were used as initial guess values. 
    \item We then limit the dataset to the data smaller than 20° of phase angle, fixing $\omega$, $g$, and $\bar{\theta}$ to the best-fit values previously obtained. We searched for solutions within the boundaries: $B_{sh,0}$ = \{0.0,3.0\}, $h_{sh}$ = \{0.0,0.15\}, and if we considered the CBOE: $B_{cb,0}$ = \{0.0,1.0\}, $h_{cb}$ = \{0.0,1.0\}. The uncertainties associated with the retrieved OE parameters mainly come from the different flux levels between the opposition effect datasets.
    \item We run again the minimization procedure on the entire dataset, now fixing the OE parameters, and with the other parameters free within $\omega$ = \{0.01,0.3\}, $g$ = \{-1.0,1.0\} (or in the case of the use of the 2T-HG: $b$ = \{0.0,1.0\} and $c$ = \{-1.0,1.0\}), $\bar{\theta}$ = \{5°,50°\}.
\end{enumerate}
For the KS model, as the parameter are not strongly correlated with each other, the process was more straightforward, with a fit on the entire dataset, searching for solutions within the following boundaries: $A_N$ = \{0.0,0.2\}, $\nu_1$ = \{0.0,10.0\}, $\nu_2$ = \{0.0,5.0\}, $m$ = \{0.0,10.0\}, and depending on the disk function $c_l$ = \{0.0°,5.0°\} or $k$ = \{0.0°,5.0°\}.\newline
It is noteworthy that, unfortunately, the Mars Express observations only cover the sub-Martian hemisphere of Deimos. Indeed, after filtering the pixels with the above conditions, we found that Deimos observations cover a surface with a latitude between 50°S and 75°N, and a longitude between 135°W and 90°E (Fig. \ref{fig:SRC_ill_conditions}).

\begin{table*}[]
    \centering
    \caption{Hapke parameters found from the global SRC disk-resolved analysis using different versions of the Hapke model, and different disk functions for the Kaasailanen-Shkuratov model.}
    \resizebox{\textwidth}{!}{
    \begin{tabular}{ccccccccccc}
    \hline
    \hline
       Model & $\omega$ & $g$ or $b$ & $c$ & $B_{sh,0}$\tablefootmark{$\dagger$} & $h_{sh}$\tablefootmark{$\dagger$} & $\bar{\theta}$ [deg] & $B_{cb,0}$\tablefootmark{$\dagger$} & $h_{cb}$\tablefootmark{$\dagger$} & Porosity & RMS\\
    \hline
       H2012-1THG\tablefootmark{a} & 0.068 $\pm$ 0.001 & -0.275 $\pm$ 0.009 & -- & 2.14 $\pm$ 0.14 & 0.065 $\pm$ 0.004 & 19.4 $\pm$ 0.1 & -- & -- & 85.7\% & 0.00439\\
       H2012-2THG\tablefootmark{b} & 0.075 $\pm$ 0.003 & 0.29 $\pm$ 0.01 & 0.63 $\pm$ 0.09 & 2.14 $\pm$ 0.14 & 0.065 $\pm$ 0.004 & 21.2 $\pm$ 0.6 & -- & -- & 85.7\% & 0.00438\\
       H2012-1THG-CBOE\tablefootmark{c} & 0.067 $\pm$ 0.002 & -0.275 $\pm$ 0.008 & -- & 2.14 $\pm$ 0.14 & 0.065 $\pm$ 0.004 & 18.6 $\pm$ 0.6 & 0.69 $\pm$ 0.20 & 0.29 $\pm$ 0.15 & 85.7\% & 0.00506\\
       H2012-2THG-CBOE\tablefootmark{d} & 0.072 $\pm$ 0.003 & 0.28 $\pm$ 0.02 & 0.75 $\pm$ 0.10 & 2.14 $\pm$ 0.14 & 0.065 $\pm$ 0.004 & 19.8 $\pm$ 0.9 & 0.69 $\pm$ 0.20 & 0.29 $\pm$ 0.15 & 85.7\% & 0.00507\\
    \hline
    \hline
       Model & $A_N$ & $\nu_1$ & $\nu_2$ & $m$ & $c_l$ & $k$ & & & & RMS \\
    \hline
       KS1\tablefootmark{e} & 0.080 $\pm$ 0.001 & 10.0$^{*}$ & 1.07 $\pm$ 0.03 & 1.73 $\pm$ 0.02 & 0.88 $\pm$ 0.04 & -- & & & & 0.00439\\
       KS2\tablefootmark{f} & 0.080 $\pm$ 0.001 & 10.0$^{*}$ & 1.10 $\pm$ 0.01 & 1.72 $\pm$ 0.01 & -- & 0.54 $\pm$ 0.03 & & & & 0.00442\\
       KS3\tablefootmark{g} & 0.077 $\pm$ 0.001 & 8.88 $\pm$ 0.06 & 0.91 $\pm$ 0.02 & 1.35 $\pm$ 0.10 & -- & -- & & & & 0.00494\\
    \hline
    \end{tabular}
    }
    \label{tab:comp_disk_resolved_analysis_model}
    \tablefoot{\\
    $\dagger$ The OE parameters were obtained with a first inversion using only data at phase angles smaller than 20°, and fixed afterwards.\\
    \tablefoottext{a}{H2012-1THG: Hapke 2012 model with porosity factor, single-term Henyey Greenstein phase function}\\
    \tablefoottext{b}{H2012-2THG: Hapke 2012 model with porosity factor, double-term Henyey Greenstein phase function}\\
    \tablefoottext{c}{H2012-1THG-CBOE: Hapke 2012 model with porosity factor, single-term Henyey Greenstein phase function, coherent-backscattering opposition effect}\\
    \tablefoottext{d}{H2012-2THG-CBOE: Hapke 2012 model with porosity factor, double-term Henyey Greenstein phase function, coherent-backscattering opposition effect}\\
    \tablefoottext{e}{KS1: Kaasalainen-Shkuratov model with double-exponential phase function, and McEwen disk function}\\
    \tablefoottext{f}{KS2: Kaasalainen-Shkuratov model with double-exponential phase function, and Minnaert disk function}\\
    \tablefoottext{g}{KS3: Kaasalainen-Shkuratov model with double-exponential phase function, and Akimov disk function}
    }
\end{table*}

\begin{table}[]
    \centering
    \caption{Physical quantities (normal albedo $A_n$, and hemispherical albedo $A_h$) derived from the global SRC disk-resolved analysis using different versions of the Hapke model.}
    \resizebox{0.5\textwidth}{!}{
    \begin{tabular}{ccc}
    \hline
    \hline
       Model & $A_n$ & $A_h$\\
    \hline
       H2012-1THG & 0.078 $\pm$ 0.006 & 0.0150 $\pm$ 0.0005\\
       H2012-2THG & 0.077 $\pm$ 0.009 & 0.0161 $\pm$ 0.0007\\
       H2012-1THG-CBOE & 0.13 $\pm$ 0.04 & 0.0145 $\pm$ 0.0005\\
       H2012-2THG-CBOE & 0.12 $\pm$ 0.05 & 0.0156 $\pm$ 0.0006\\
    \hline
    \end{tabular}
    }
    \label{tab:disk_resolved_derived_quantities}
\end{table}

\subsubsection{Surface reflectance variations with the SRC}
\begin{figure*}
    \centering
     \begin{subfigure}[b]{0.49\textwidth}
         \centering
         \resizebox{\hsize}{!}{\includegraphics[width=\textwidth]{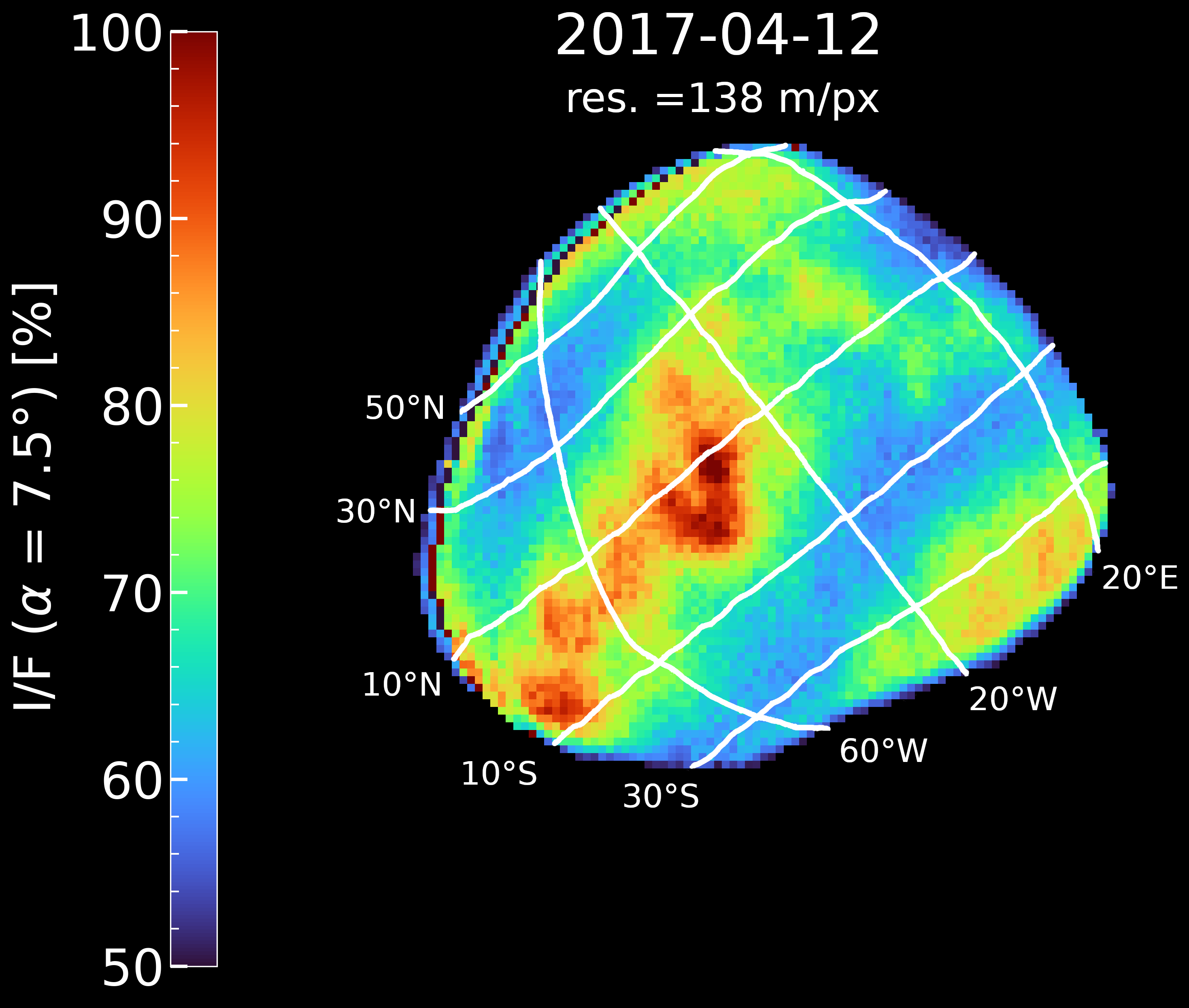}}
         \caption{}
         \label{fig:IFcolor_deimos1}
     \end{subfigure}
     \hfill
     \begin{subfigure}[b]{0.49\textwidth}
         \centering
         \resizebox{\hsize}{!}{\includegraphics[width=\textwidth]{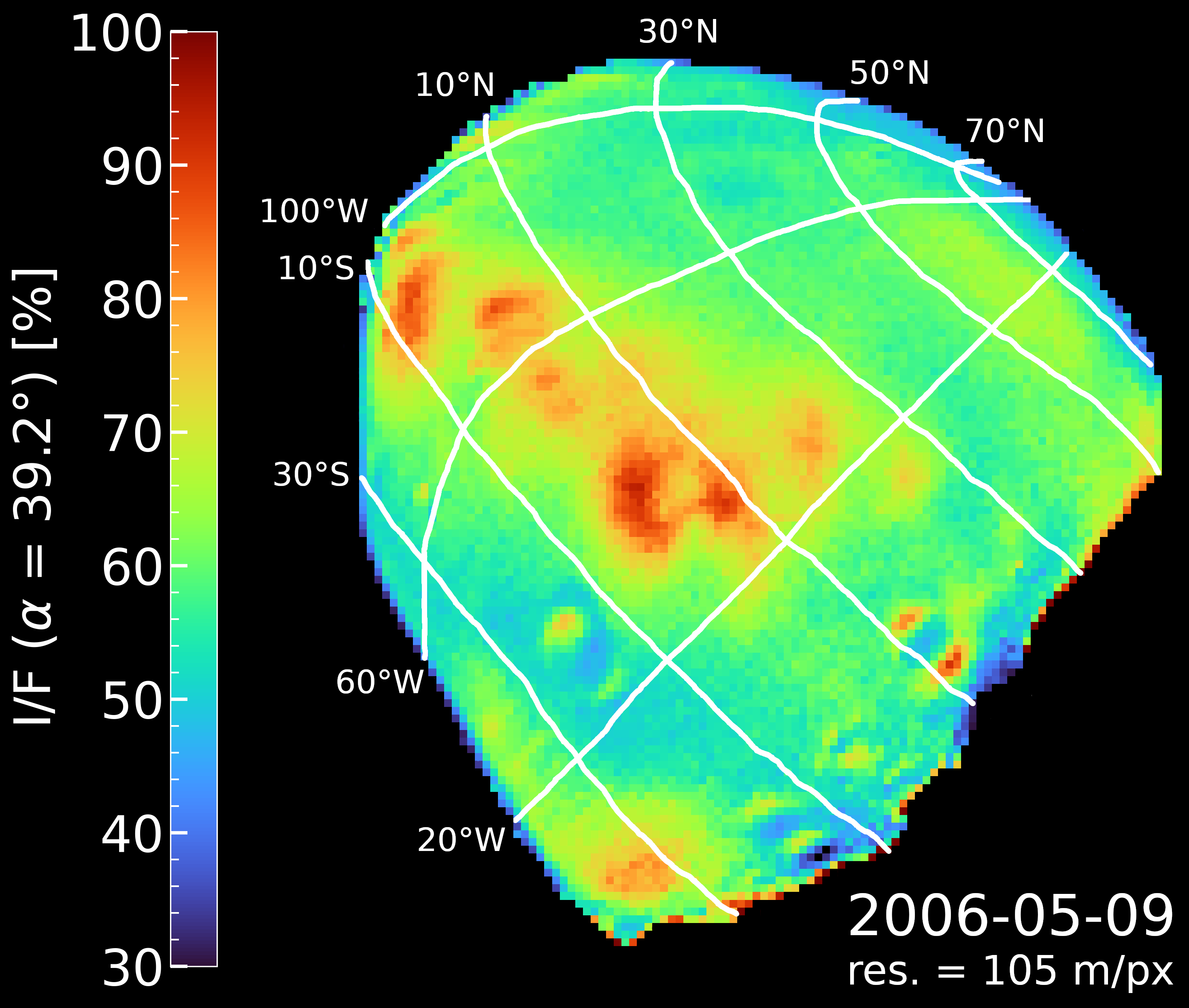}}
         \caption{}
         \label{fig:IFcolor_deimos2}
     \end{subfigure}
     \caption{Relative I/F images of Deimos from the SRC at different phase angles and different spatial resolutions.}
     \label{fig:IF_colordeimos_resolved}
\end{figure*}

We observed that the ridge around the equatorial plane is the brighter location of Deimos (visible in the used image because the image covers only a fraction of Deimos, Fig. \ref{fig:HRSC_deimos_disk-resolved} and Fig. \ref{fig:IF_colordeimos_resolved}). Except for this feature, the north and south of the Voltaire and Swift craters also appear to be brighter than the average surface brightness of Deimos but less than the ridge. The average reflectance of the surface is between 0.06-0.07, whereas the north and south of the Voltaire and Swift craters are around 0.08, and the mean for the ridge is 0.085-0.09. Hence, the ridge appears to be $\sim$35\% brighter than the average reflectance of Deimos. The region of the ridge with the highest reflectance is 58\% brighter. It is also worth noting that Deimos is homogeneous except for these features. The Deimos craters (e.g., Voltaire and Swift) also do not show a particular increase of reflectance (color variation) on the crater rims as seen previously in the case of Phobos for the different craters. \par 

\subsubsection{Results of the disk-resolved Hapke model}
The goal of the use of the photometric models, and in particular the Hapke model, is also to describe the surface regolith properties. Even if it has been shown that directly linking physical properties to Hapke parameters is difficult \citep{Shepard_2007, Helfenstein_2011}, it is still useful to compare with the parameters of other objects to constrain the surface properties.

\begin{figure*}
    \centering
     \begin{subfigure}[b]{0.49\textwidth}
         \centering
         \resizebox{\hsize}{!}{\includegraphics[width=\textwidth]{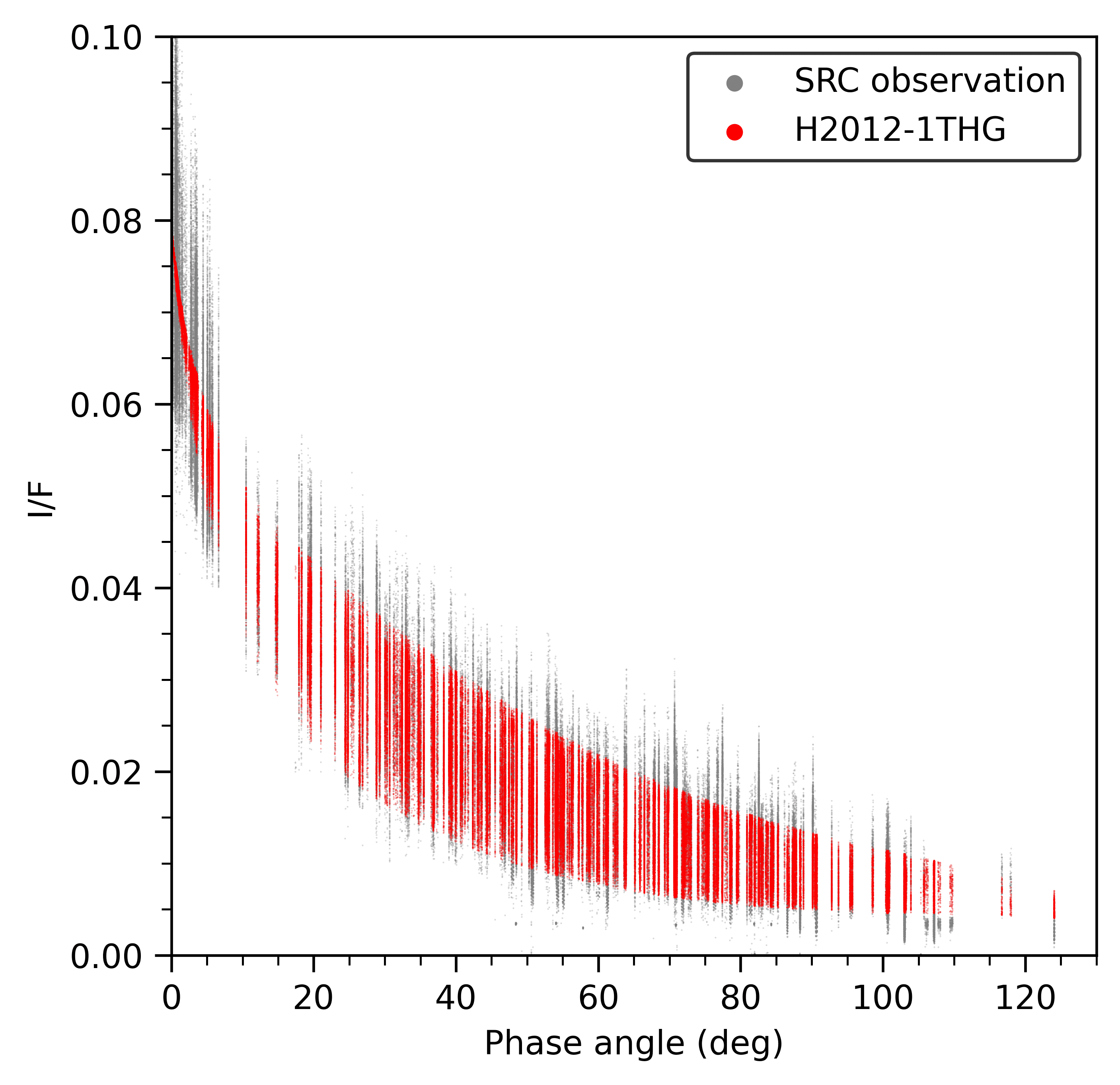}}
         \caption{}
         \label{fig:disk_resolved_2THG_1}
     \end{subfigure}
     \hfill
     \begin{subfigure}[b]{0.49\textwidth}
         \centering
         \resizebox{\hsize}{!}{\includegraphics[width=\textwidth]{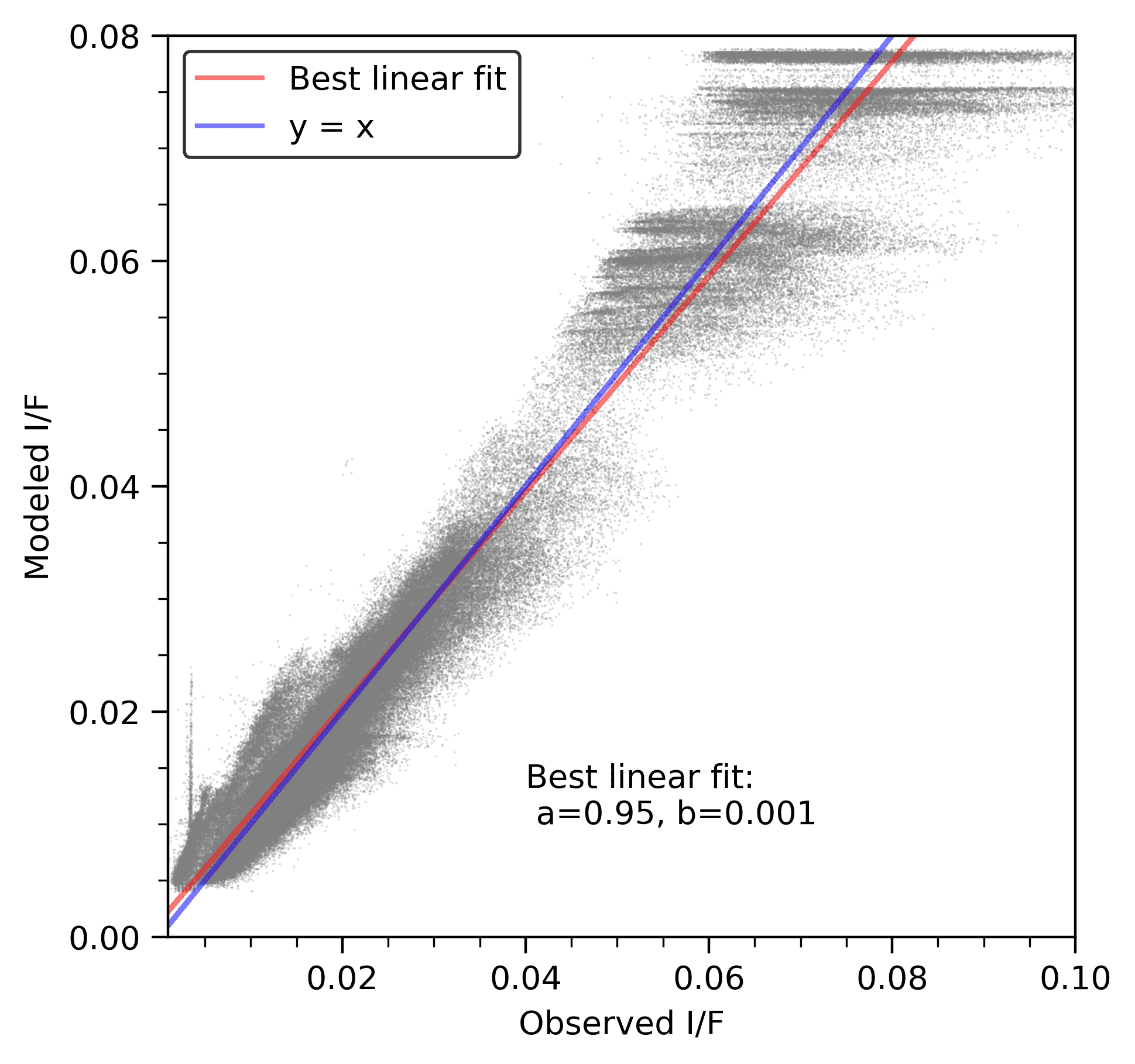}}
         \caption{}
         \label{fig:disk_resolved_2THG_2}
     \end{subfigure}
     \caption{Disk-resolved photometry (gray points) of Deimos obtained with the SRC. Each point represents the radiance factor of an element of the surface at a given illumination condition (incidence, emission, phase). The illumination conditions were retrieved from SPICE simulations, using the shape model from \cite{Ernst_2023}. The modeled data, using the Hapke H2012-2THG model, are shown as red dots.}
     \label{fig:disk_resolved_2THG}
\end{figure*}

\paragraph{Whole observed surface analysis}
The first analysis was performed considering the data for the entire surface of Deimos. The results of the disk-resolved Hapke global fit are shown in Table \ref{tab:comp_disk_resolved_analysis_model}. An exemple of Hapke disk-resolved model fit with the H2012-2THG is shown in Fig. \ref{fig:disk_resolved_2THG}.\newline
Because of our inversion procedure, the SHOE parameters are kept fixed for the four versions of the Hapke model. The CBOE parameters are also fixed when coherent-backscattering opposition surge is considered. The inversion with the different versions of the model shows that the best-fit is achieved when using the 2T-HG phase function and no CBOE. The SHOE intensity ($B_{sh,0}$) shows that Deimos is strongly affected by the opposition surge, with an important increase of the reflectance at small phase angle. The SHOE parameters show that Deimos surface layer is probably made of opaque grains in the form of fractal aggregates, producing a highly porous (86\%) optical first layer (on the first tens of µm). The derived SSA and the asymmetry parameter $g$ show that the surface of Deimos is dark and predominantly backscatters the light ($g < 0$). We found that the addition of the CBOE does not increase the quality of the fit. The CBOE is associated with coherent interferences which arise near-opposition ($<$2°). The Deimos opposition surge can be fully modeled only with the SHOE contribution. Our photometric analysis revealed that the CBOE is negligible on Deimos, as it is expected for dark surfaces \citep{Shevchenko_2012}. However, from the CBOE parameters derived, we can still try to tentatively describe the grain structures of the surface. In particular, $h_{cb}$ is linked to the transport mean free path \citep{Hapke_2012}:
\begin{equation}
    h_{cb} = \frac{\lambda}{4 \pi \Lambda}
\end{equation}
where $\lambda$ is the wavelength of observation, and $\Lambda$ the transport mean free path. Considering the derived value of the half-width of the CBOE, we obtained an estimated mean free path of 180 nm. The value can give a tentative indication about the distribution of the scattering centers. The results for our data would likely indicate that a photon arriving at the surface of Deimos will travel about 1/3 of its wavelength before being scattered. This may be due to the high micro-porosity, such as cracks or pores \citep{Coulson_2007, Consolmagno_2008, Noguchi_2015, Ostrowski_2019}, or micro-structures notably due to space-weathering with bubbles and vesicles \citep{Noguchi_2023, Rubino_2024} on this type of grain. 

\begin{figure}
\centering
  \resizebox{\hsize}{!}{\includegraphics{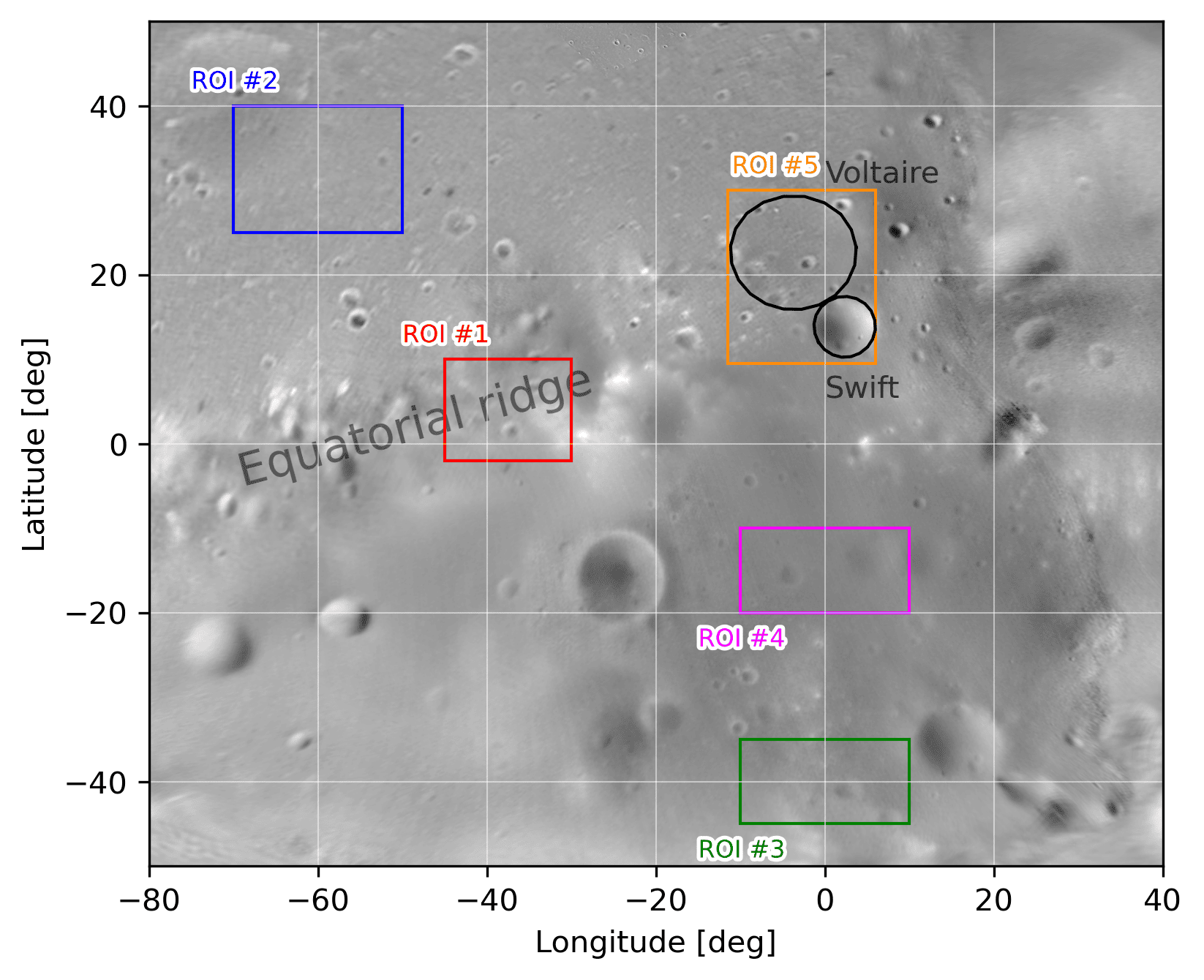}}
  \caption{Positions of the selected ROIs on a global map in cylindrical projection \citep{Stooke_2015}. ROI \#1 was taken on the equatorial ridge, ROI \#3 on the bright region in the south of Deimos, ROI \#2 and ROI \#4 on dark regions respectively at the north of the ridge, and at the north of ROI \#3. The ROI \#5 was defined to include the two craters Voltaire and Swift.}
  \label{fig:ROIs}
\end{figure}

\paragraph{Regional analysis}
We performed a disk-resolved photometric analysis using the H2012-1THG on five regions of interest (ROIs), selected for the presence of geological features or for their specific brightness behavior. The selected regions are indicated in Table \ref{tab:ROIs} and Fig. \ref{fig:ROIs}. \par
The SSA exhibits clearly different values for the different ROIs. While dark ROIs have an SSA of approximately 6.2\%, the other brighter regions show an SSA between 6.8\% and 8.2\%.  The highest SSA is found for the streamers on the equatorial ridge. The ROIs with the two craters, Voltaire and Swift, have a higher SSA than many surface regions of Deimos. However, with our spatial resolution, it does not seem to be correlated with the crater rims as it is for Phobos (around the Stickney crater for example), but more with an overall higher reflectance in this region. This region shows reflectance heterogeneities and the bright ridge may be a contributing factor.\newline
We also looked at the variations of the backscattering across the surface (asymmetry parameter $g$). No specific correlation is visible between the different defined ROIs, either based on the reflectance of the surface, on the geological features, or on the locations (e.g., northern/southern hemisphere). \newline
Looking at the opposition effect parameters, the half-width of the SHOE varies between 0.066 and 0.073 for the five ROIs. The highest $h_{sh}$ are associated with the dark regions, while the slightly smaller half-widths of the SHOE are correlated with the brighter regions of the surface. On the other hand, the amplitude of the SHOE is correlated with the reflectance of the ROIs. The bright areas exhibit a higher $B_{sh,0}$ parameter (2.12), whereas the dark regions have a slightly smaller amplitude with a value of 1.9-2.0. A higher amplitude of the SHOE is generally associated with complex-shaped grains creating a high porosity, as well as the presence of opaque materials at the surface. Therefore, the bright regions, including the equatorial ridge, would likely be composed of more porous materials. \newline
The roughness parameter $\bar{\theta}$ was found to strongly vary depending on the ROIs. The ridge exhibits a $\bar{\theta}$ of 21°, relatively similar to the average value of this parameter. The two selected dark regions have a smaller roughness parameter of approximately 15°. The bright region on the southern hemisphere (ROI \#3) has a particularly small $\bar{\theta}$ of 8°, implying a tentatively particularly smooth area on Deimos, maybe due to the presence of particularly fine grains. On the other hand, the craters region (ROI \#5) exhibits a high roughness ($\bar{\theta}=28$°), which may be linked to the presence of the two craters in this ROI.

\paragraph{Photometric parameters map}
In order to account for all regional differences and the potential heterogeneity of the Deimos surface, we considered the fitting of the surface not only with a single global function, but also with a spatially variable function. In order to perform this analysis, the data were first binned into several longitude/latitude bins. This was achieved from the SRC images and their associated co-registered geometrical images. A grid of longitude and latitude with bins of size 5° $\times$ 5° was created, and each pixel of the images in the bins was associated based on their longitude and latitude. The longitude/latitude binning size was chosen to have enough spatial resolution but also to ensure enough data are present in each individual bin. We ignored (i) pixels with incidence or emission larger than 70°, (ii) bins containing less than 50 data points, (iii) bins with phase angle at least covering between 2° and 90°. The result for each bin was a phase curve on which we could apply photometric models. For each bin, we inverted the Hapke model to retrieve the Hapke parameters and the associated quantities, such as the normal albedo and the hemispherical albedo. The RMS map is also computed, based on the residuals of each bin. This process was performed for both H2012-1THG and H2012-2THG models. We did not consider the CBOE, as we found previously its contribution to be negligible. \par
\begin{figure*}
\resizebox{\hsize}{!}{\includegraphics{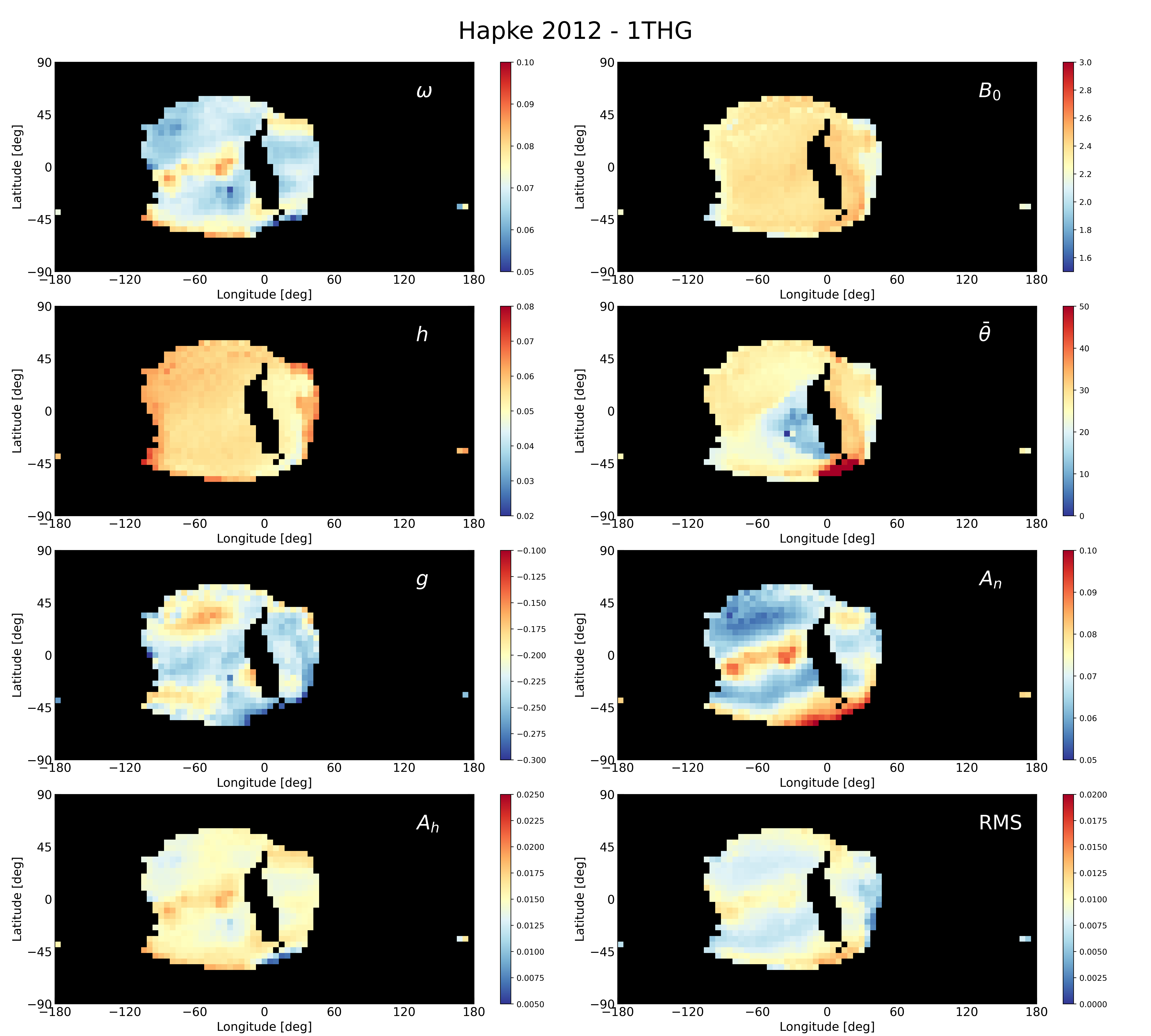}}
\caption{Hapke parameter maps (H2012-1THG) and related quantities (normal albedo $A_n$ and hemispherical albedo $A_h$) derived with the SRC observations at 650 nm. The RMS error map appears to be linked with the position of the ridge. Black areas represent regions with no sufficient data. The data are projected on the map using the equirectangular projection.}
\label{fig:HRSC_params_map_1THG}
\end{figure*}

For both H2012-1THG (Fig. \ref{fig:HRSC_params_map_1THG}) and H2012-2THG (Fig. \ref{fig:HRSC_params_map_2THG}), we found a clear correlation between the position of the equatorial ridge and the SSA. Except for this region, the SSA is almost constant across the surface. In contrast to the slight variations observed with the ROI analysis, we did not find modifications of the SHOE parameters, either in amplitude or in half-width. This is likely linked to a smaller amount of data due to the smaller regions sampled in each bin. Surprisingly, the roughness parameter $\bar{\theta}$ is smaller ($<$ 20°) in the southeast region of the ridge, while the other parts of the surface exhibit a roughness of $\sim$24°.  For the H2012-2THG, it is particularly difficult to build a map of the $b$ and $c$ parameters. In the case of the H2012-1THG, we computed the asymmetry parameter $g$ for each bin. The region to the north of the equatorial ridge has the higher $g$ values ($g \simeq -0.17$). The region at the south of the ridge also exhibits a value higher than the average value ($g \simeq -0.21$). The ridge itself has the same $g$ as the average surface and does not show specific behavior. From the Hapke parameters, we can derive again related quantities such as the hemispherical albedo and the normal albedo. The two albedos are correlated with the position of the ridge. In particular, $A_n$ is 25\% higher (than the global normal albedo) on the ridge, while the variations of $A_h$ are smaller ($\sim$12\%). \par
\begin{figure*}
\resizebox{\hsize}{!}{\includegraphics{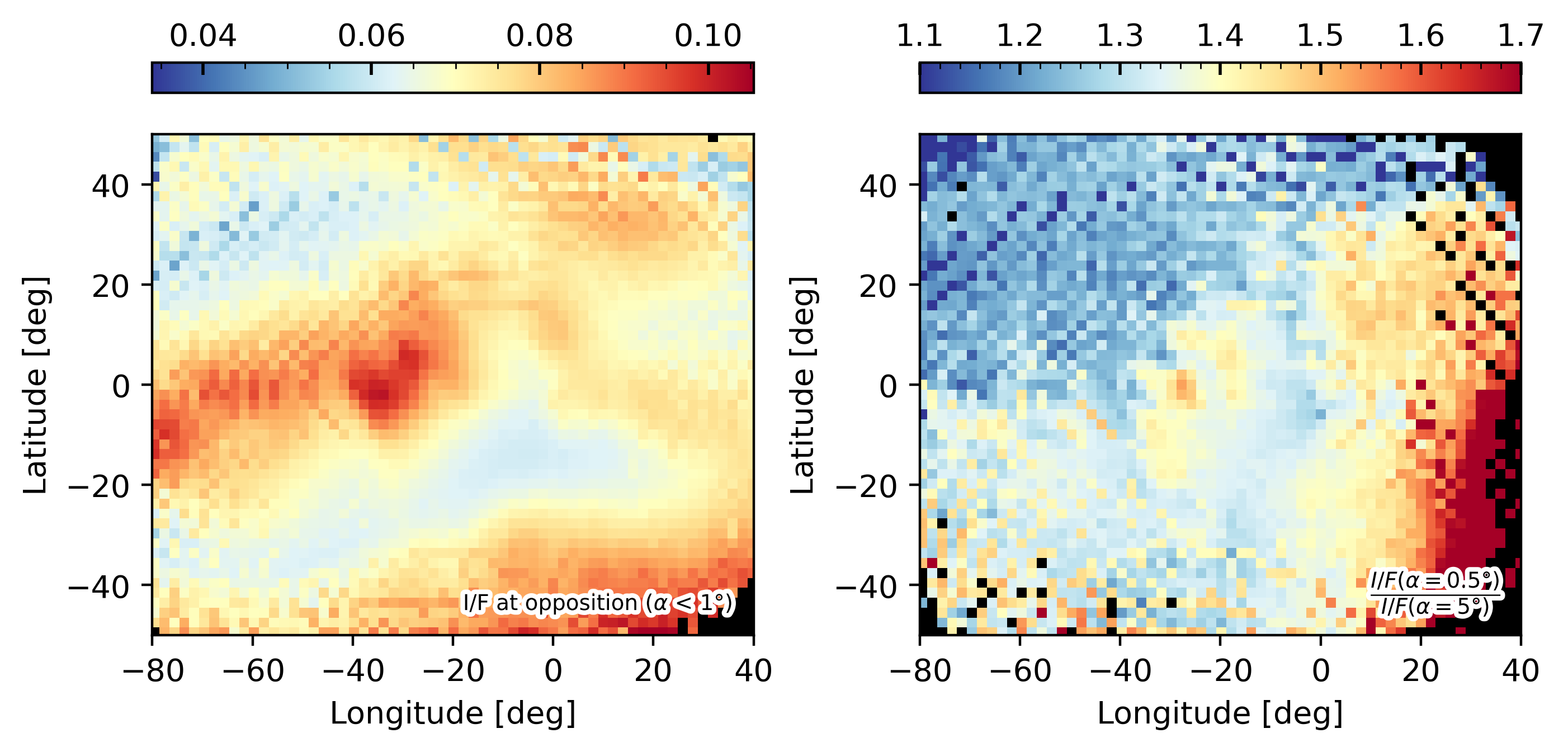}}
\caption{(a) Spatial variations of the radiance factor at phase angle smaller than 1°. (b) Spatial variation of the opposition surge: ratio of the radiance factor at a phase angle of 0.5° and the radiance factor at a phase angle of 5°. The data were binned into a 2°$\times$ 2° longitude-latitude grid. When several values are present for a given bin, the median of the values is considered. The black pixels correspond to regions with no data. The data are projected on the map using the equirectangular projection.}
\label{fig:IF_opposition_maps}
\end{figure*}

We confirmed the results obtained using the Hapke parameters by computing a map of the radiance factor at the opposition (i.e., for a phase angle of less than 1°; Fig. \ref{fig:IF_opposition_maps}a). Additionally, we derived the phase ratio (i.e., the ratio of the radiance factor at a phase angle of 0.5° to that at a phase angle of 5°), allowing to visualize the spatial variations of the opposition surge (Fig. \ref{fig:IF_opposition_maps}b). The spatial variations of the radiance factor are consistent with the distributions of the single-scattering albedo and the normal albedos from the Hapke model. The opposition surge shows no clear spatial variations. In particular, no variations are linked with geological features on Deimos, such as the equatorial ridge or craters.

\begin{table*}[]
    \centering
    \caption{Hapke parameters found from the disk-resolved analysis of five regions of interest on Deimos, using the H2012-1THG model.}
    \resizebox{\textwidth}{!}{
    \begin{tabular}{cccccccccc}
    \hline
    \hline
       ROIs\tablefootmark{a} & Regions & Latitude & Longitude & $\omega$ & $g$ & $B_{sh,0}$ & $h_{sh}$ & $\bar{\theta}$ [deg] & Porosity\\
    \hline
       \#1 & Ridge & 2°S - 10°N & 30°W - 45°W & 0.082 $\pm$ 0.001 & -0.276 $\pm$ 0.001 & 2.12 $\pm$ 0.01 & 0.067 $\pm$ 0.001 & 21.3 $\pm$ 0.2 & 85.3\%\\
       \#2 & Dark & 25°N - 40°N & 50°W - 70°W & 0.061 $\pm$ 0.001 & -0.247 $\pm$ 0.001 & 1.90 $\pm$ 0.01 & 0.073 $\pm$ 0.001 & 14.5 $\pm$ 0.3 & 84.2\%\\
       \#3 & Bright & 35°S - 45°S & 10°W - 10°E & 0.068 $\pm$ 0.001 & -0.308 $\pm$ 0.001 & 2.12 $\pm$ 0.01 & 0.067 $\pm$ 0.001 & 7.9 $\pm$ 0.5 & 85.4\%\\
       \#4 & Dark & 10°S - 20°N & 10°W - 10°E & 0.063 $\pm$ 0.001 & -0.241 $\pm$ 0.001 & 1.98 $\pm$ 0.01 & 0.069 $\pm$ 0.001 & 16.1 $\pm$ 0.2 & 85.0\%\\
       \#5 & Craters & 9.5°N - 30°N & 11.5°W - 6°E & 0.070 $\pm$ 0.001 & -0.265 $\pm$ 0.001 & 2.13 $\pm$ 0.01 & 0.066 $\pm$ 0.001 & 28.1 $\pm$ 0.2 & 85.5\%\\
    \hline
    \end{tabular}
    }
    \label{tab:ROIs}
    \tablefoot{
    \tablefoottext{a}{The location of the different ROIs is also indicated in Fig. \ref{fig:ROIs}.}
    }
\end{table*}

\subsubsection{Results of the Kaasalainen-Shkuratov models}
The Kaasailanen-Shkuratov model has also been widely used in the literature to perform photometric correction of the data (e.g., \citealt{Domingue_2016, Hasselmann_2016, Schroder_2018, Li_2021, Zou_2021, Golish_2021, Golish_2021b, Yokota_2022}). However, the possibility to link the parameters with the physical properties of the surface is less evident than the Hapke model, and the relative comparison with other studies is also more complicated, as many combinations of disk function and phase function can be made. Therefore, it is rare that different studies adopt the same model. The choice of disk function and phase function is generally made based on the type of surface and the dataset (e.g., data at opposition). The primary goal of this analysis was to provide additional photometric model parameters for the preliminary photometric correction of the MMX instruments. The photometric correction allows to provide a standardization to similar illumination conditions. Generally, the standard geometry of observations is $(i,e,\alpha) = (30°, 0°, 30°)$. The results of the disk-resolved KS global fit are shown in Table \ref{tab:comp_disk_resolved_analysis_model}. The best fits were achieved with the McEwen and the Minnaert disk functions. The three KS models defined in this work demonstrate a relatively similar normal albedo, ranging from 0.077 to 0.080. We checked the quality of the photometric correction for the various models performed in this study. To this end, two images were selected, obtained at different phase angles (14.9° and 40.6°) and exhibiting an overlapping region (Fig. \ref{fig:photometric_correction_1}). The data with incidence and emission larger than 70° was removed. We computed the radiance factor profile of these two images for a given latitude (Fig. \ref{fig:photometric_correction_2}). While the Akimov model is adequate for moderate illumination conditions, it poorly reproduces the radiance factor for high incidence or emission angles. The models proposed in this work, along with the associated parameter provide already a reasonable and satisfactory photometric correction. The improvement of this photometric correction could be achieved by calculating maps for each parameter, rather than utilising the global fit parameter as previously performed.

\subsubsection{Local spectro-photometry}
\begin{figure*}
\centering
\resizebox{\hsize}{!}{\includegraphics{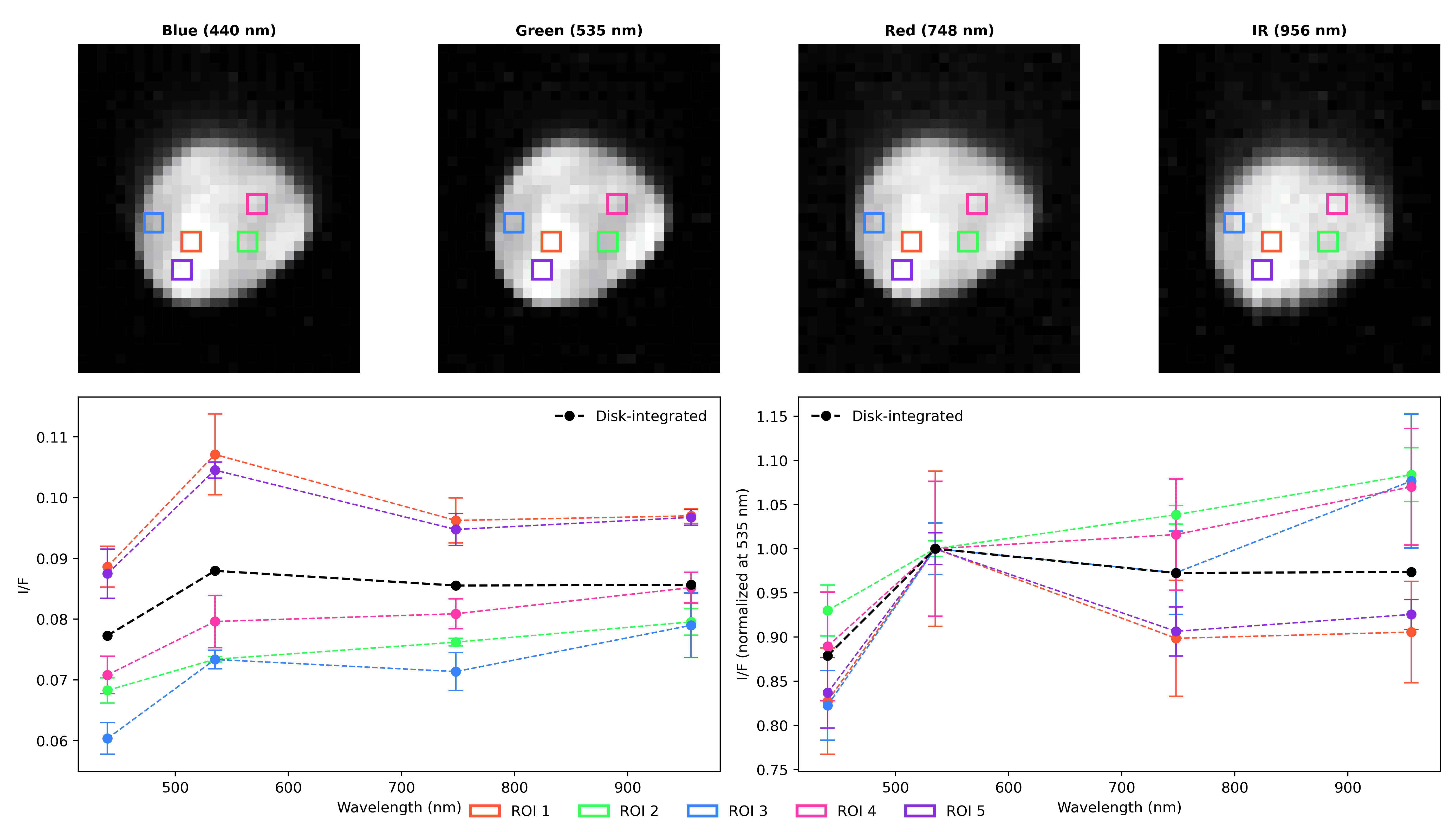}}
\caption{Deimos spectro-photometry using the four filters of the HRSC. Each ROI has a size of 2 pixels by 2 pixels. The red and purple ROIs (ROI 1 and ROI 5) correspond to the equatorial ridge. The green ROI 2 was chosen because it corresponds to the darkest region of the surface as visible in the HRSC images. The two other ROIs are representative of the average surface. The bottom left panel presents the spectra associated with each of the ROIs. The black spectrum is derived from the disk-integrated analysis using aperture photometry. The bottom right panel corresponds to the spectra for each ROI, normalized at the green filter central wavelength (i.e., 535 nm). Note that the red and IR radiance factors are lower than expected for Deimos; this is linked to an absolute calibration factor problem with these filters \citep{McCord_2007}.}
\label{fig:Deimos_spectrophometry}
\end{figure*}

Previous spectroscopic observations were insufficiently spatially resolved to derive spectra in different regions of Deimos \citep{Fraeman_2012}. To study the spectroscopic properties of Deimos, we extracted the pixels (390-800 m/px) on the geometrically corrected images, corresponding to several regions including the equatorial ridge; computed the mean flux from these pixels, and then compared them with the disk-integrated HRSC Deimos spectrum (Fig. \ref{fig:Deimos_spectrophometry}). Note that the ROIs defined for this spectroscopic analysis differ from those in Fig. \ref{fig:ROIs} due to to the different spatial resolution and surface coverage between the SRC and HRSC cameras. \newline 
We confirmed that the ridge is brighter than all other regions on the Mars-near side of Deimos in all four filters of the HRSC camera. Fig. \ref{fig:Deimos_spectrophometry} (top) presents the different ROIs selected, the corresponding spectrum is shown on the bottom left panel, and the bottom right presents the normalized spectra at 550 nm. The darkest regions located at the south of the equatorial ridge appear to be in some images redder than the average surface. The ridge region is bluer than the average surface. It is noteworthy that the spectral slopes on Deimos with this analysis are much lower than expected, ranging from 2.5\%.(100 nm)$^{-1}$ for the ridge to 4.4\%.(100 nm)$^{-1}$ for the other regions on average. The typical Deimos spectral slope in the same wavelength range (blue-IR) was found from spectra in the literature to be 10.8\%.(100 nm)$^{-1}$ \citep{Fraeman_2012, Takir_2021}. However, considering only the blue-green slope the HRSC data are consistent with previous observations, with a spectral slope of approximately 4\%.(100 nm)$^{-1}$. The observed difference for the red and IR filters is due to an issue with the absolute calibration factor provided in the header of each HRSC image. This problem was already reported in \cite{McCord_2007}. However, \cite{McCord_2007} noticed an unexpected increase in the flux in the red filter and a decrease in the IR filter. The red filter behavior is different from what we observed, with a decrease in the flux for both red and IR filters. The reason for this discrepancy is not yet understood. However, the relative modifications of the spectral slopes between the ridge and the other regions are not affected by this issue. This problem was already visible for Phobos observations with slightly lower flux compared to other observations (but still within error bars), but the observations of Deimos show a difference between the HRSC radiance factor and the CRISM radiance factor \citep{Fraeman_2012} at the same wavelength of a factor of two. 

\section{Discussion}
\subsection{Photometric comparison between Deimos and Phobos}
Within uncertainties, the amplitude of the SHOE $B_{sh,0}$ (2.14 $\pm$ 0.14) is similar to the one derived for Phobos ($2.28 \pm 0.03$, \citealt{Fornasier_2024}). The half-width of the SHOE $h_{sh}$ (0.065 $\pm$ 0.004) is slightly higher than the one derived for Phobos. Both Martian moons are mainly backscattering ($g = -0.275$ for Deimos and $g = -0.267$ for Phobos in the green filter). The roughness parameter $\bar{\theta}$ is slightly smaller for Deimos ($\bar{\theta} = 19-21$°) compared to Phobos ($\bar{\theta} = 24$°), which would likely indicate that the surface of Deimos is smoother than Phobos. The interested reader is referred to \cite{Fornasier_2024} for a comprehensive comparison of the Hapke parameters with other objects of the Solar system. \par
\begin{figure}
\centering
\resizebox{\hsize}{!}{\includegraphics{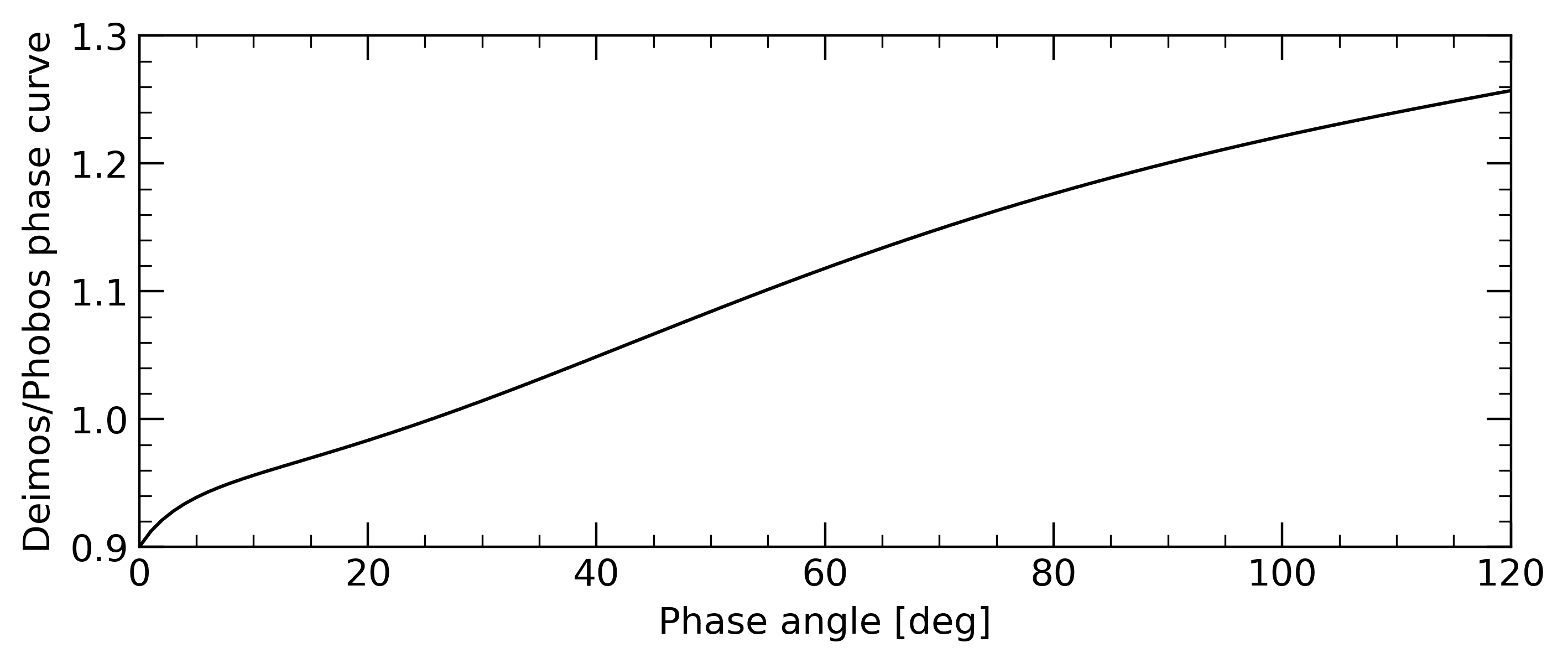}}
\caption{Ratio of the disk-integrated Deimos phase curve over the Phobos curve. The two phase curves were derived from the SRC camera and obtained through Hapke modeling. The Phobos phase curve was plotted from Hapke parameters from \cite{Fornasier_2024}.}
\label{fig:deimos_vs_phobos_phase_curve}
\end{figure}

From the Hapke parameters obtained in this work and by the study of \cite{Fornasier_2024}, we can compare the phase curves of the two Martian moons  (Fig. \ref{fig:deimos_vs_phobos_phase_curve}). Deimos is slightly darker than Phobos up to 25° of phase angle, and becomes rapidly brighter after. At a phase angle of 120°, the Deimos surface is 25\% brighter than Phobos at the same phase angle. \par
From our photometric analysis, we derived a porosity of the top-layer surface of Deimos of approximately 85\%. Radar observations of Deimos have also shown the presence of high-porosity surface ($\sim$60\%) while Phobos should have a smaller porosity of approximately 40\% \citep{Busch_2007}. However, these differences in porosity were not observed between this study on Deimos and the photometric study on Phobos by \cite{Fornasier_2024}. The differences between radar and visible light are that the two techniques probe different parts of the surface. While our photometric study describes the first microns of the surface, the radar observations on a body like Deimos can give information on the first two meters (e.g., \citealt{Kamoun_2014}). Therefore, the comparison between this study, and the work by \cite{Fornasier_2024}, and \cite{Busch_2007}, indicates that while the two Martian moons are highly porous at the upppermost (photometrically-active) layer of approximately 85-90\% and that both become more compact with depth, Phobos porosity decreases rapidly with depth (only the first tens of microns are very porous on Phobos) but the surface of Deimos would have a thicker porous dust layer.\par
\begin{figure}
\resizebox{\hsize}{!}{\includegraphics{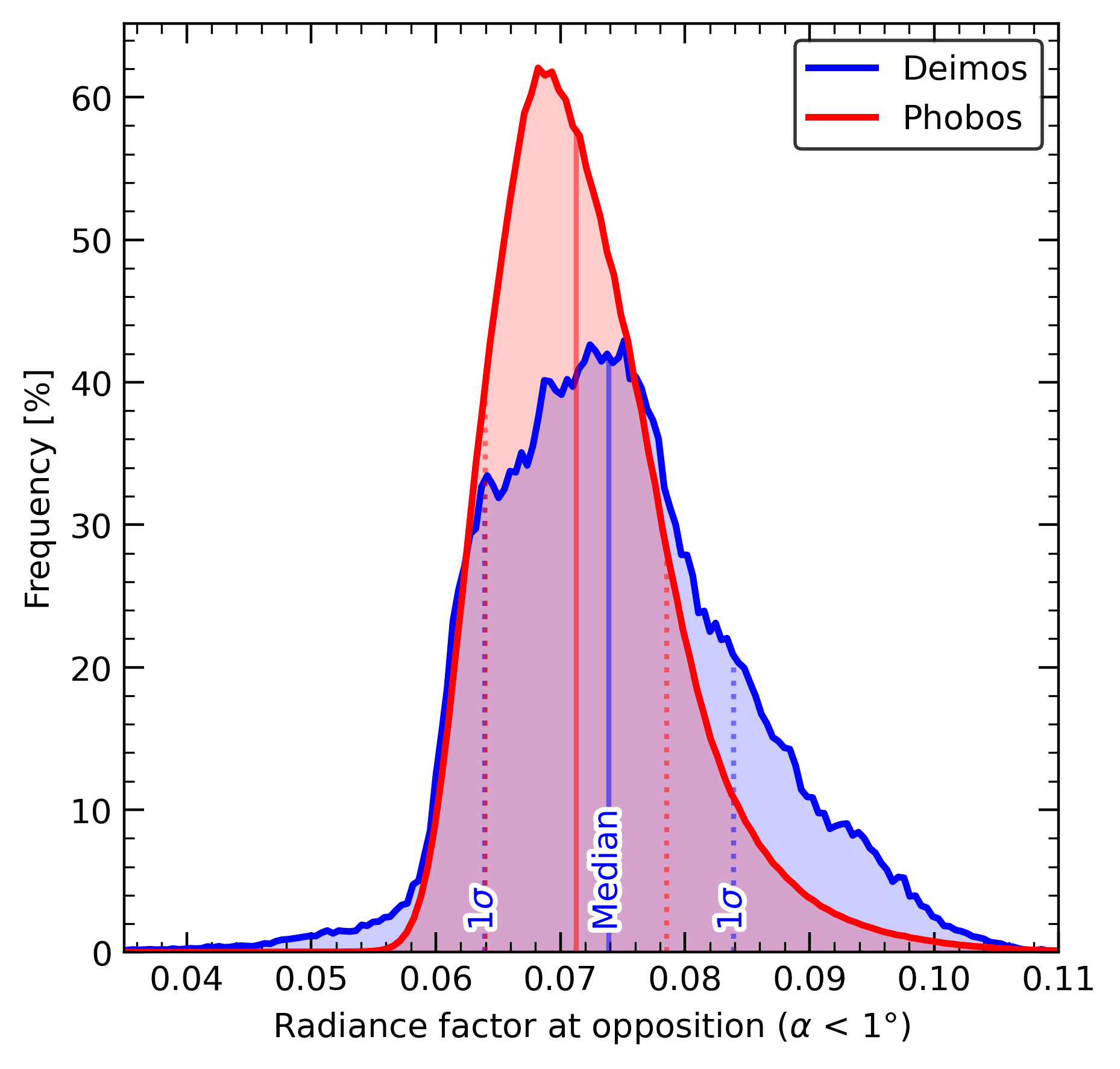}}
\caption{Frequency of the radiance factor values for Deimos (blue) and Phobos (red), considering only pixels with phase angle smaller than 1°. The vertical solid lines represent the median of the radiance factor values, and the vertical dotted lines represent the $\pm$1$\sigma$ values. The Phobos radiance factor values distribution is computed using 32 million pixels, and the Deimos one using 270\,000 pixels.}
\label{fig:hist_phobos_deimos}
\end{figure}

In order to better observe the global surface variations of the radiance factor, we plotted the distribution of the radiance factors for both Phobos and Deimos (Fig. \ref{fig:hist_phobos_deimos}). In order to avoid the contribution of the shadows in the data, we selected only data obtained at opposition ($\alpha < 1$°), and only selected pixels with incidence and emission angles smaller than 75°. The Phobos data comes from the same instrument, using data from \cite{Fornasier_2024}. Note that for the two Martian moons, only the sub-Martian hemisphere was observed by Mars Express. We clearly observed that the radiance factor of Phobos is much more peaked, while for Deimos it is more spread out over a wide range of values. The median value of Phobos is 0.071 with a standard deviation of 0.007. Therefore, approximately 95\% of the radiance factor values are contained within the range 0.050 - 0.092. For Deimos, we found a slightly higher median value of 0.074, with a larger standard deviation of 0.010. The Deimos radiance factor is then contained between 0.044 and 0.104 for 95\% of the values. While the Phobos surface exhibits more radiance factor variations across the surface, Deimos has a broader distribution, indicating that, if the surface is clearly more homogeneous in comparison to Phobos, the bright regions of Deimos cover a larger area (with respect to the size of the object). Additionally, this broader distribution is also more asymmetric than the one of Phobos, with the same sharp rise, but a very gradual tail at larger radiance factor values.\newline
On Phobos, the bright regions were found to be mainly related to crater ejecta \citep{Fornasier_2024}. On Deimos, we found that this is mostly linked with the global scale topography, as already observed by \cite{Thomas_1996}. \par
Despite some slight photometric variations between the two Martian moons, Phobos and Deimos are extremely similar and share many photometric properties, including the opposition effect. 

\subsection{A blue unit on Deimos}
Based on the HRSC images in four different color filters, we tentatively identified the presence of a blue unit on Deimos. We then confirmed the results of the analysis of the High Resolution Imaging Science Experiment (HiRISE) onboard Mars Reconnaissance Orbiter (MRO) by \cite{Thomas_2011}, which created a blue-red slope image showing the blue behavior of the ridge/streamers and of the southern region below the ridge. No spectrometer (e.g., ISM onboard Phobos2, OMEGA onboard MEx, or CRISM onboard MRO) was sufficiently spatially resolved to decipher this blue unit on Deimos. Using the blue and IR filters, we found a spectral slope variation of about 50\% between the two units on Deimos. This variation is more important than the one derived by \cite{Thomas_2011} of the order of 10\%. However, it should be kept in mind that our analyses are partially hindered by the calibration issue for the HRSC red and IR filters. \par
The Deimos blue unit is very similar to the blue unit on Phobos, with a higher reflectance and a less red spectral slope. On Phobos, the blue unit is brighter by 40-50\% than the average surface \citep{Fornasier_2024}. The spectral slope in the VNIR is decreasing by approximately 60\% for the Phobos blue unit \citep{Wargnier_2025}. On Deimos, we showed that the blue unit has a radiance factor higher by 35-58\% than the average surface, and a spectral slope smaller than 50\% compared to the Deimos red unit. The variations between the blue and red units on the two Martian moons are extremely similar. \par
The physical reason for the presence of this blue unit on Deimos is intriguing, as for the one on Phobos. On Phobos, the blue unit has been mainly explained by the fact that it corresponds to fresher materials, as shown by the evidence that the Phobos blue unit is mostly located on crater rims, and would therefore correspond to fresh crater ejecta (e.g., \citealt{Murchie_1991, Murchie_1996}). Because the Limtoc crater is a crater inside the huge crater Stickney, Limtoc is younger than Stickney, and Limtoc should be bluer than its surroundings. However, this is not the case, and this theory has been considered to be unlikely by \cite{Thomas_2011}. Another hypothesis has been made based on the evidence that the Phobos blue unit is not only located on crater rims but also, for example, on the floor of the Stickney crater \citep{Thomas_2011}. Based on these observations, \cite{Basilevsky_2014} suggested that Phobos may be made of a mixture of blocks of red and blue units distributed across the surface. A recent study also suggests that the Phobos blue unit may be the result of the modification of the texture of the surface (e.g., porosity, roughness, grain size) \citep{Wargnier_2024b}. \par
The photometric behavior of the bright streamers (corresponding to the Deimos blue unit) located on the equatorial ridge may then be explained in different ways. One important property is that the ridge obviously has a higher geopotential height as shown in \cite{Thomas_1979, Thomas_1996}, resulting in a downslope trend from the top of the ridge. \cite{Thomas_1979} suggested that the streamers would correspond to a very thin layer of downslope materials that is likely fresher than their surrounding. \cite{Thomas_1996} also proposed that the streamers are more likely linked to fresher areas because they did not find modifications of the phase function with albedo. Based on our analysis, in particular the albedo variations and the slight modifications of the opposition effect, we suggest that the Deimos blue unit is more likely caused by textural modification, such as the presence of fine grain or a higher porosity. The finer grain of carbonaceous chondrite has already shown a brighter and less red behavior consistent with the blue unit observed in this work \citep{French_1988}. In particular, the blue unit (i.e., the streamers in our images) will be possibly linked with segregation of the grains by the topography trending downslope from the top of the ridge. Other investigations will be needed to fully characterize this blue unit and to help identify if other occurrences of bright and blue areas can be found. The analysis of the data obtained by the Hera mission, as well as the future observation of the MMX mission will be pivotal to advance our knowledge of the Deimos surface.

\section{Conclusions}
We have analyzed photometric observations of Deimos obtained from 2004 to 2025 by the HRSC and SRC cameras onboard Mars Express. The first part was dedicated to the absolute calibration of the SRC using mutual events observations with Jupiter and stars, as well as observations made with the HRSC absolutely calibrated blue, green, red, and IR filters. The SRC dataset is unique for its coverage in time, phase angle (0.06 - 120°), number of images ($>$3000), and with one of the best spatial resolutions so far for observations of Deimos. We summarize here the main findings of our work:
\begin{enumerate}
    \item[-] Deimos has a strong opposition surge ($B_{sh,0}= 2.14 \pm 0.14$ and $h_{sh} = 0.065 \pm 0.004$) due to shadow-hiding. The contribution of the coherent-backscattering process is negligible. The opposition effect of Deimos is very similar to the one on Phobos \citep{Fornasier_2024}.
    \item[-] The albedo of Deimos is slightly higher (7.4\%) than that of Phobos (7.1\%), but it still has a very dark surface. A comparison of the phase curves of Phobos and Deimos in the four HRSC filters shows that Deimos is slightly brighter, particularly in the red and IR filters, but very similar in the blue and green filters.
    \item[-] The distribution of the radiance factor for Phobos and Deimos shows a very similar behavior with a sharp rise for the low radiance factor and a gradual decrease for the brightest regions. The distribution is broader on Deimos. 
    \item[-] The top-layer surface probe by our analysis shows that Phobos and Deimos are extremely similar on the first few microns, composed of opaque materials, complex-shaped grains or fractal aggregates forming a porous layer (86\%). However, considering the results of radar observations, Deimos is probably much more porous even in the first meters. This would indicate that Deimos is covered by a thick dust layer, consistent with the hypothesis of an impact cratering at the origin of the concave depression at the south pole of Deimos, producing a regolith mantle \citep{Thomas_1989}. This is also in agreement with the fact that we do not observe specific photometric behavior around craters, indicating that the craters are relatively old, partially filled by regolith, and therefore show the same albedo. 
    \item[-] We observed the presence of a blue unit on Deimos, confirming the result obtained by \cite{Thomas_2011}. This Deimos blue unit is located at the equatorial ridge from 10°W to 90°W, on structures called streamers. We did not find any occurrence of other blue unit regions on the Mars near-side of Deimos observed by the HRSC/SRC. Similarly to Phobos, the Deimos blue unit consists of a brighter region with a less red spectral slope. The red-blue unit variations are similar between Phobos and Deimos. On Deimos, the blue unit is 35\% to 58\% brighter than the average surface and exhibits a decrease of the spectral slope by approximately 50\%.

\end{enumerate}
The extreme similarity between the photometric properties of the surface of Phobos and Deimos, including the opposition effect, as well as the tentative presence of a blue unit on Deimos (in a similar way as Phobos), would likely indicate that Phobos may have follow a similar history and that the Martian moons come from a unique parent body. Therefore, Phobos and Deimos may originate either from the disruption of a unique asteroid or bilobated body as hypothetized by \cite{Fornasier_2024}, \cite{Kegerreis_2025}, and \cite{Wargnier_2025}; or from accretion after a giant impact (e.g., \citealt{Craddock_2011, Rosenblatt_2016, Hyodo_2017, Canup_2018}). The possibility of two different captured asteroids appears much more unlikely.\par
This study will also be useful in the context of the Martian Moon eXploration mission \citep{Kuramoto_2022}, for the autoexposure algorithm of the MIRS spectrometer \citep{Barucci_2021}, and for the first photometric correction which will be applied to the MIRS data, as well as for the TENGOO and OROCHI cameras \citep{Kameda_2021}. The observations by the different instruments will help to confirm the presence of this blue unit on Deimos, clarify its origins, and try to find other regions where the blue unit is visible. \par
Although the SRC calibration factor we derived was not subjected to extensive tests on Mars data, we believe it could also be beneficial for the Mars community and anyone interested in examining photometric phenomena occurring on the surface of Mars. 

\section*{Data availability}
The data cubes prepared for this study, which contain the HRSC or SRC images as well as all the co-registered geometric backplanes (incidence, emission, phase, latitude and longitude), can be downloaded from the following Zenodo repository: \url{https://doi.org/10.5281/zenodo.17081101}

\begin{acknowledgements}
    We acknowledge the Centre National d'Etudes Spatiales (CNES) for their financial support. We are grateful to the anonymous reviewer for providing insightful comments and suggestions that improved the paper. We thank the ESA Planetary Science Archive for space mission data procurement and the principal investigator of the HRSC instrument, G. Neukum (Freie Universität, Berlin, Germany). We warmly thank H. U. Keller, R. Ziese, and A. Pommerol for the useful discussions and advice on the absolute calibration. We are grateful to R. Sultana for the valuable discussions on the radar observations. This research has made use of the SIMBAD database, operated at CDS, Strasbourg, France. This work has made use of data from the European Space Agency (ESA) mission {\it Gaia} (\url{https://www.cosmos.esa.int/gaia}), processed by the {\it Gaia} Data Processing and Analysis Consortium (DPAC, \url{https://www.cosmos.esa.int/web/gaia/dpac/consortium}). Funding for the DPAC has been provided by national institutions, in particular the institutions participating in the {\it Gaia} Multilateral Agreement.
\end{acknowledgements}

\bibliographystyle{aa}
\bibliography{references.bib}

\begin{appendix}
\section{Additional information about the absolute calibration of the SRC}
This section presents additional information for the absolute calibration of the SRC.

\subsection{Spectra of stars used for calibration}
\begin{figure*}
\resizebox{\hsize}{!}{\includegraphics{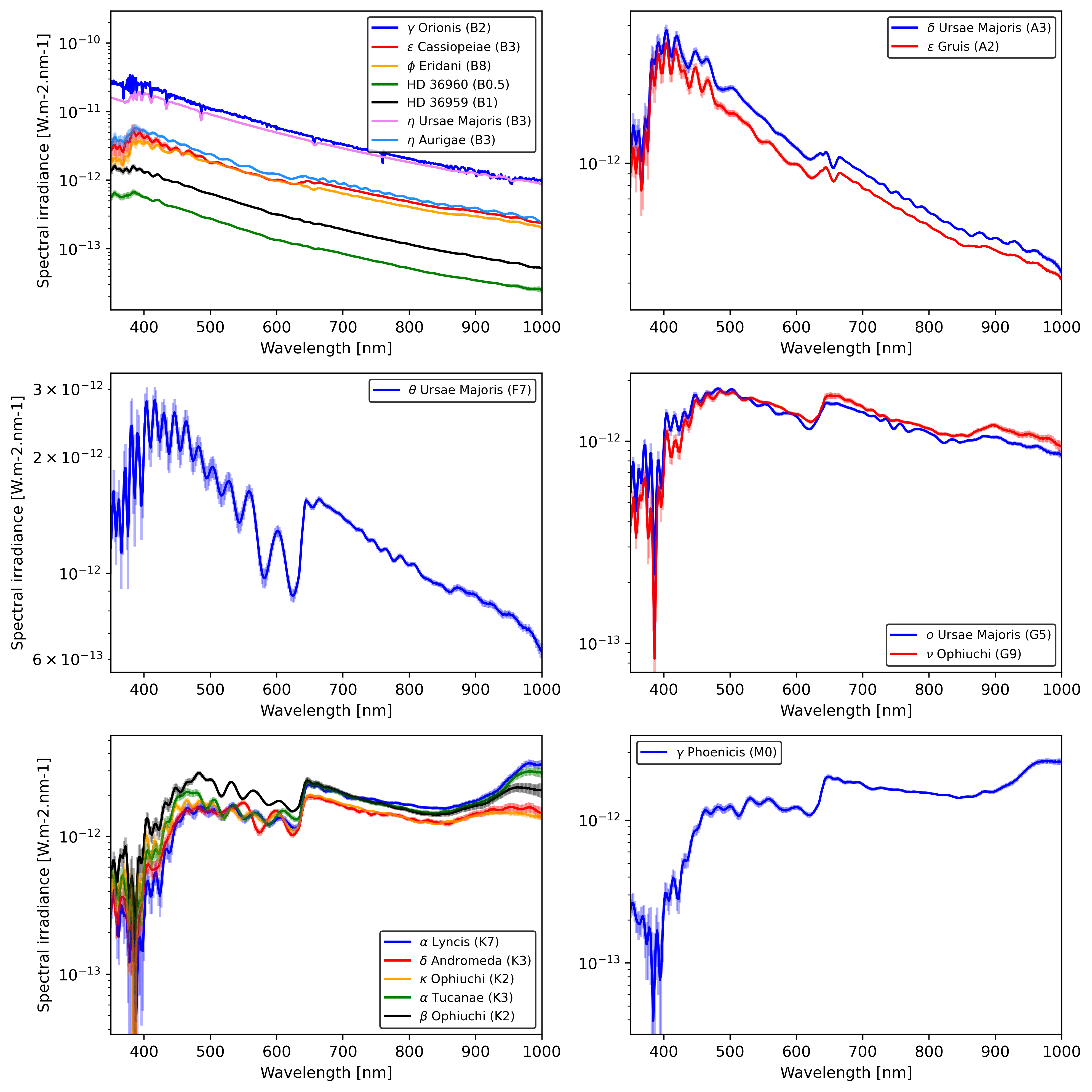}}
\caption{Spectra of the stars used for absolute calibration of the SRC. Spectra come from the Gaia DR3 catalog, except for $\gamma$ Orionis where the spectrum was taken from \cite{Krisciunas_2017}.}
\label{fig:spec_stars}
\end{figure*}

For each identified star, we retrieved the spectrum when available in the Gaia DR3 catalog. The exception is $\gamma$ Orionis for which the spectrum was obtained in \cite{Krisciunas_2017}. Fig. \ref{fig:spec_stars} presents the spectra used for each star classified by spectral types (from B- to M-types).

\subsection{HRSC vs. SRC calibration}
For the cross-calibration between HRSC and SRC, we plotted the calibrated flux of the HRSC (in radiance factor) as a function of the SRC flux measured in DN/s, for observations made in the same conditions. We observed a linear relation between the two fluxes for the four different filters, therefore giving an absolute calibration factor.

\begin{figure*}
\resizebox{\hsize}{!}{\includegraphics{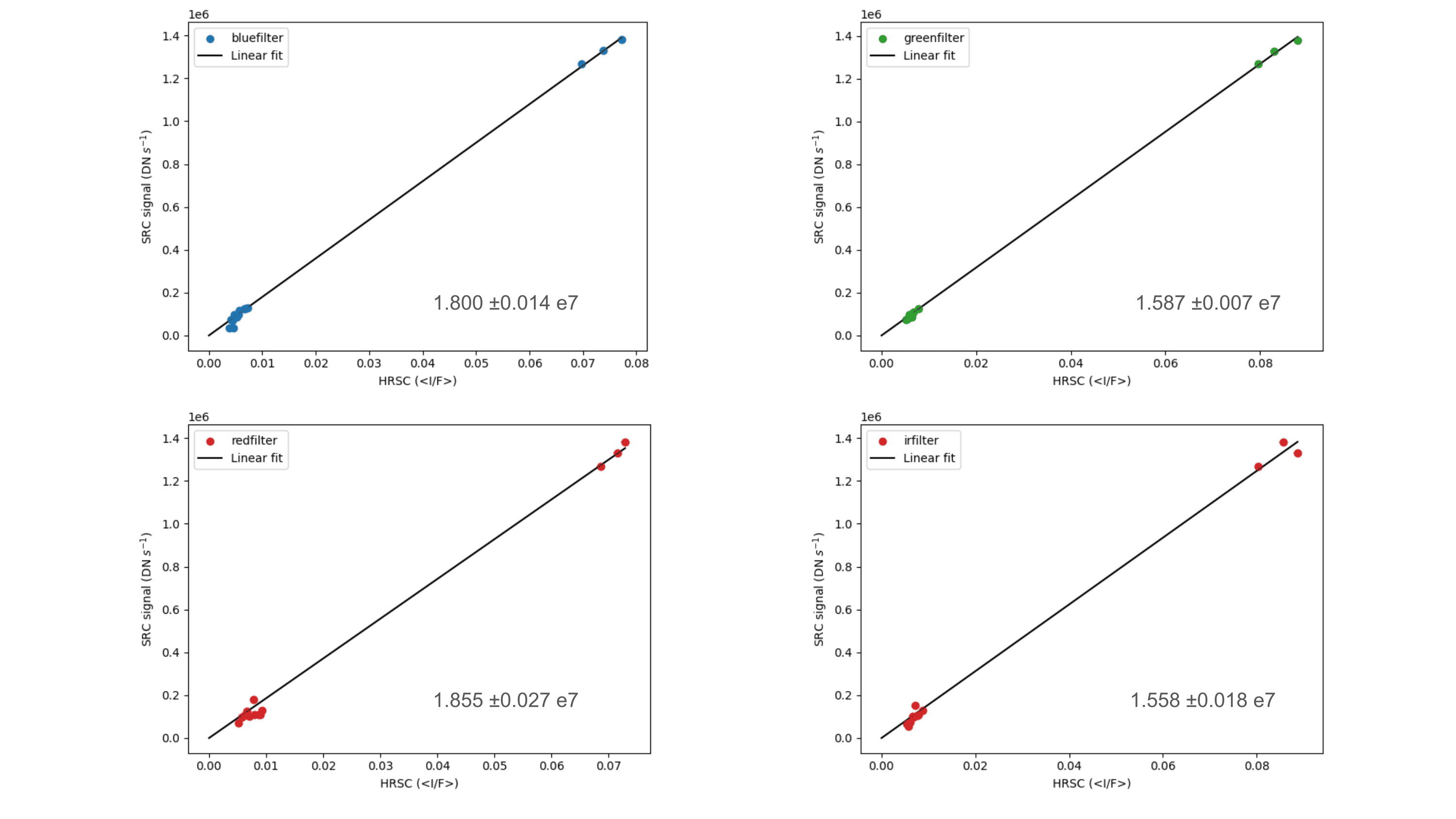}}
\caption{HRSC signal measured in the four filters vs. SRC signal for Deimos observations.}
\label{fig:HRSC_vs_SRC}
\end{figure*}

\subsection{Calibration factors for each object}
\begin{figure}
\resizebox{\hsize}{!}{\includegraphics{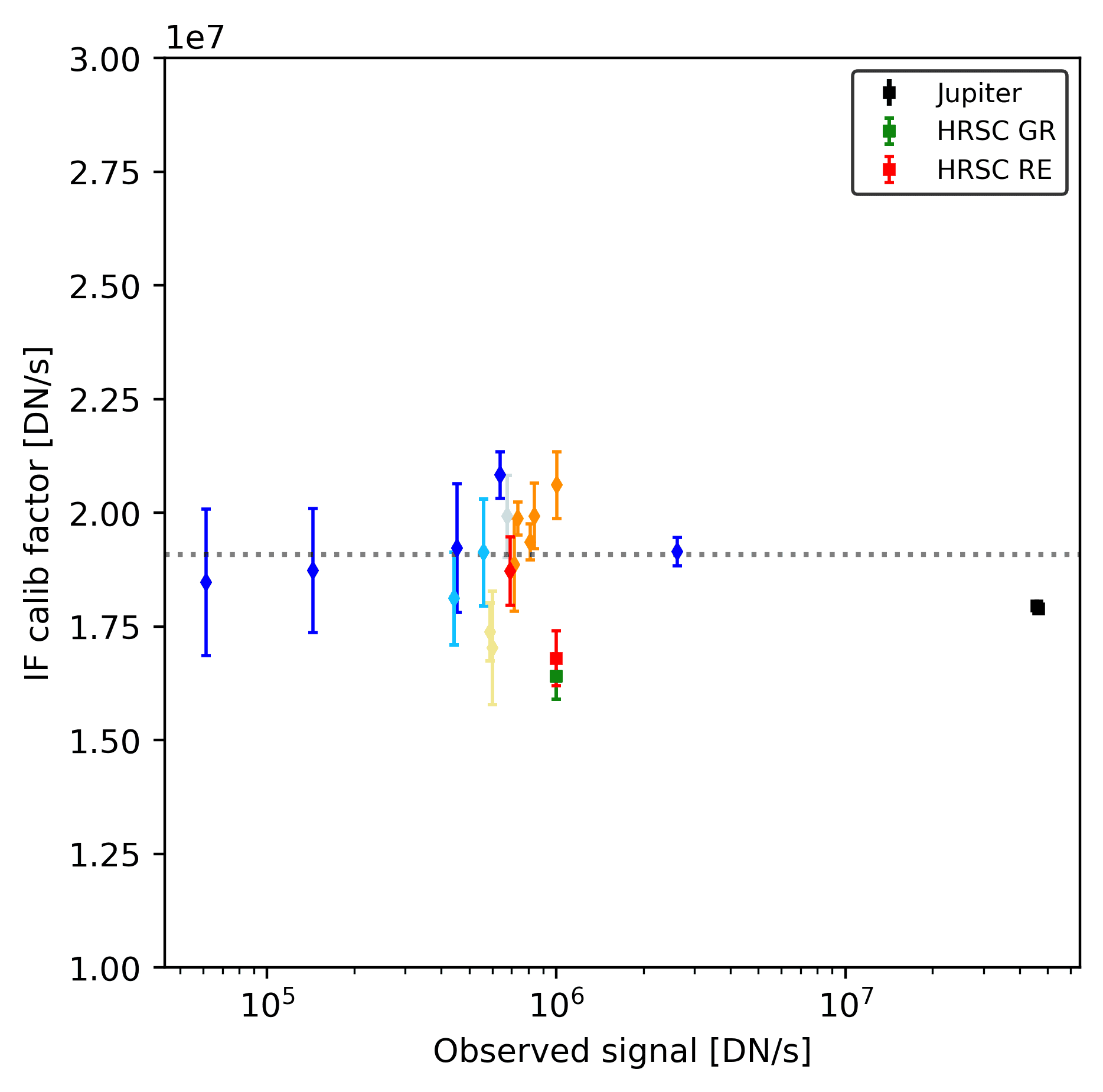}}
\caption{I/F calibration factor for each object/method as a function of the observed signal. The black squares represent Jupiter data; the red and green squares represent, respectively the I/F calibration factor derived from the HRSC red filter data and the HRSC green filter data. The diamond shapes correspond to the stars used in this study, with the different colors representing the different spectral types. The horizontal dotted line is the average I/F calibration factor derived from stars only.}
\label{fig:IF_calib}
\end{figure}

We present the I/F calibration factor derived for each object/method in Fig. \ref{fig:IF_calib}. We chose to give an equal weight to each method used to derive the absolute calibration factors. Therefore, we computed the average calibration factors given by the 17 star observations, the average I/F from the two Jupiter observations, and the average calibration factor from the HRSC observations in the green and red filters. The results of the three calculations are given in Table \ref{tab:absolute_cal_results}.

\begin{table}[]
    \centering
    \caption{Results of the absolute calibration factors of the SRC for each method/object.}
    \begin{tabular}{ccc}
    \hline
    \hline
    Method/object & $A_{I/F}$\tablefootmark{a} \\
    \hline
    Jupiter & (1.79 $\pm$ 0.03) $\times$ 10$^{7}$ \\
    Jupiter & (1.79 $\pm$ 0.03) $\times$ 10$^{7}$ \\
    Stars & (1.90 $\pm$ 0.10) $\times$ 10$^{7}$ \\
    HRSC BL & (1.80 $\pm$ 0.01) $\times$ 10$^{7}$ \\
    HRSC GR & (1.59 $\pm$ 0.01) $\times$ 10$^{7}$ \\
    HRSC RE & (1.85 $\pm$ 0.03) $\times$ 10$^{7}$ \\
    HRSC IR & (1.56 $\pm$ 0.02) $\times$ 10$^{7}$ \\
    \hline
    \end{tabular}
    \label{tab:absolute_cal_results}
    \tablefoot{
    \tablefoottext{a}{$A_{I/F}$ is given in DN/s.}
    }
\end{table}

\section{Photometric models} 
\subsection{Hapke model} \label{appendix:hapke}
\subsubsection{Disk-resolved Hapke model}
We used the Hapke IMSA model \citep{Hapke_2012} with the porosity correction, shadowing function, and shadow-hiding opposition effects:
\begin{multline}
    \frac{I}{F} = K \frac{\omega}{4} \frac{\mu_{0,e}}{\mu_{0,e}+\mu_{e}} S(i,e,\alpha,\bar{\theta}) \left[1+B_{cb}(\alpha, B_{cb,0}, h_{cb})\right] \\ \times \left\{P_{hg}(\alpha,g) \left[1+B_{sh}(\alpha, B_{sh,0}, h_{sh})\right] + M\left(\frac{\mu_{0,e}}{K}, \frac{\mu_e}{K}, \omega \right)\right\}
\end{multline}
where $\mu_0$ and $\mu$ are respectively the cosine of the effective incidence and emergence angles, and $\omega$ is the single-scattering albedo. \newline
The $B_{sh}$ function describes the shadow-hiding opposition effect (SHOE):
\begin{equation}
    B_{sh}(\alpha,B_{sh,0},h_{sh}) = \frac{B_{sh,0}}{1 + \frac{\tan \alpha/2}{h_{sh}}}
\end{equation}
where $B_{sh,0}$ is the amplitude of the SHOE, and $h_{sh}$ is the half-width of the SHOE.\newline
The $B_{cb}$ function describes the coherent-backscattering opposition effect (CBOE):
\begin{equation}
    B_{cb}(\alpha,B_{cb,0},h_{cb}) = \frac{B_{cb,0}}{1 + 1.42K} \frac{1}{\left(1+\frac{\tan{\alpha/2}}{h_{cb}}\right)^{2˛}} \left[1 + \frac{1 - e^{-1.42K\frac{\tan{\alpha/2}}{h_{cb}}}}{\frac{\tan{\alpha/2}}{h_{cb}}} \right]
\end{equation}
where $B_{cb,0}$ is the amplitude of the CBOE, and $h_{cb}$ is the half-width of the CBOE.\newline
M is the multiple scattering function given by the following equation:
\begin{equation}
    M\left(\frac{\mu_0}{K}, \frac{\mu}{K}, \omega \right) = H\left(\frac{\mu_0}{K},\omega \right) H\left(\frac{\mu}{K}, \omega \right) - 1   
\end{equation}
where $H$ is the Hapke's second-order approximation of the Chandrasekhar's function \citep{Hapke_2012}. $S$ is the shadowing function and is described in detail in \cite{Hapke_2012}.\newline
K is the porosity factor. We used the approximation from \cite{Helfenstein_2011} which makes the porosity factor dependent on the half-width of the SHOE:
\begin{equation}
    K = 1.069 + 2.109 h_{sh} + 0.577 h_{sh}^2 + 0.062 h_{sh}^3
\end{equation}
$P_{hg}$ is the Henyey-Greenstein (HG) phase function. We used both the one-term HG (1T-HG) and the two-term HG (2T-HG):
\begin{equation}
    P_{1T-HG}(\alpha,g) = \frac{1-g^2}{(1 + 2g\cos\alpha +g^2)^{3/2}}
\end{equation}
\begin{multline}
    P_{2T-HG}(\alpha,g) = \frac{1+c}{2}\frac{1-g^2}{(1 - 2g\cos\alpha +g^2)^{3/2}} \\+ \frac{1-c}{2}\frac{1-g^2}{(1 + 2g\cos\alpha +g^2)^{3/2}}
\end{multline}

\subsubsection{Disk-integrated Hapke model}
The disk-integrated Hapke model is given by:
\begin{multline}
    \frac{I}{F} = K(\alpha,\bar{\theta}) \{ \left[\frac{\omega}{8}((1 +B_{sh}(\alpha))P_{hg}(\alpha, g) -1) + \frac{r_{0}}{2}(1-r_{0})\right] \\
     \times \left(1-\sin\frac{\alpha}{2} \tan\frac{\alpha}{2} \ln \left[\cot\frac{\alpha}{4}\right]\right) + \frac{2}{3 \pi} r_{0}^{2} (\sin\alpha + (\pi - \alpha) \cos \alpha \}
\end{multline}
where:
\begin{equation}
    r_0 = \frac{1 - \sqrt{1 - \omega}}{1 + \sqrt{1 + \omega}}
\end{equation}
and:
\begin{equation}
    K(\alpha, \bar{\theta)} = \exp{\left[-32\bar{\theta} \left(\tan \bar{\theta} \tan \frac{\alpha}{2}\right)^{1/2} - 0.52\bar{\theta} \tan \bar{\theta} \tan \frac{\alpha}{2}\right]}
\end{equation}

\subsubsection{Other quantities derived from the Hapke model}
From the disk-integrated model, we derived several quantities: geometric albedo, bond albedo, phase integral, and porosity, which are related to the Hapke parameters. The geometric albedo is computed by the following expression:
\begin{equation}
    A_{p} = \frac{\omega}{8} \left[(1 + B_{sh,0}) P_{1T-HG}(\alpha=0°, g)- 1 \right] + U(\bar{\theta}, \omega) \frac{r_0}{2} \left(1 + \frac{r_0}{3}\right)
\end{equation}
where:
\begin{equation}
    U(\bar{\theta}, \omega) = 1 - (0.048\bar{\theta} + 0.0041\bar{\theta}^{2}) r_{0} - (0.33\bar{\theta}-0.0049\bar{\theta}^{2}) r_{0}^{2}
\end{equation}
We also computed the Bond albedo using:
\begin{equation}
    A_{B} = r_0 \left(1 - \frac{1 - r_{0}}{6}\right)
\end{equation}
The phase integral is then derived from the geometric and Bond albedos:
\begin{equation}
    q = \frac{A_B}{A_p}
\end{equation}
The filling factor $\phi$ is expressed by:
\begin{equation}
    h_{sh} = -0.3102 \phi^{1/3} \ln(1 - 1.209 \phi^{3/2})
\end{equation}
Therefore, to determine the porosity, we need to resolve the following equation for $\phi$:
\begin{equation}
    1 - 1.208994x^2 - e^{-3.223727 h_{sh}/x} = 0
\end{equation}
This equation is then solved with Newton's root finding algorithm, and then the porosity is computed by:
\begin{equation}
    P = 1 - \phi
\end{equation}
From the disk-resolved model, we derived also other quantities linked with the Hapke parameters, and in particular the normal and hemispherical albedos. The normal albedo is computed from the following equation:
\begin{equation}
    A_{n} = K \frac{\omega}{8} P_{hg}(0, g) \left(1 + B_{sh,0}\right) \left(1 + B_{cb,0}\right) 
\end{equation}
This equation is true for dark objects (such as Deimos), because it allows to neglect the contribution of the multiple scattering.
The hemispherical albedo is dependent on the incidence angle, and can be computed by integrating the radiance factor over the upper hemisphere $\Omega$:
\begin{align}
    A_h(i) &= \frac{1}{\pi \mu_0} \int_\Omega \frac{I}{F} \left(i, e, \alpha\right) \mu \,d\Omega \\
           &= \frac{1}{\pi \mu_0} \int_{\varphi=0}^{2 \pi} \int_{e=0}^{\pi/2} \frac{I}{F} \left(i, e, \alpha\right) \mu \sin{e} \,de \,d\varphi
\end{align}

\subsection{Kaasalainen-Shkuratov model} \label{appendix:KS}
We also performed inversion of the photometric properties using the Kaasalainen-Shkuratov (KS) model \citep{Kaasalainen_2001, Shkuratov_2011}. The KS model is simpler and more empirical than the Hapke model but has also been widely used for photometric correction of remote-sensing observations (e.g., \citealt{Domingue_2016, Hasselmann_2016, Domingue_2019, Golish_2021, Li_2021, Filacchione_2022}). The model is generally described by three decoupled terms:
\begin{equation}
    \frac{I}{F} = A_{N} f(\alpha, \lambda) D(\alpha, i, e, \lambda)
\end{equation}
where $A_N$ is the normal albedo, $f(\alpha, \lambda)$ the phase function, and $D(\alpha, i, e, \lambda)$ the disk function. Several phase functions and disk functions can be implemented. For this work, we will use a phase function that includes the modelling of the opposition effect:
\begin{equation}
    f(\alpha) = \frac{e^{-\nu_{1} \alpha} + m e^{-\nu_{2}\alpha}}{1 + m}
\end{equation}
where $\nu_1$ and $m$ are associated with the width and the amplitude of the opposition effect (SHOE and CBOE), respectively, and $\nu_2$ describes the surface roughness. \newline
We considered three disk functions which are the Lommel-Seeliger-Lambert ($D_{LSL}$), the Minnaert ($D_{M}$), and the Akimov ($D_{A}$) functions. The McEwen is a combination of the Lommel-Seeliger, incorporating the contribution of the Lambert correction ($cos(i)$):
\begin{equation}
    D_{LSL} = c_{l} \frac{2 \mu_{0}}{\mu_{0}+\mu} + (1-c_{l})\mu_{0}
\end{equation}
where $c_l$ corresponds to the fraction of the Lommel-Seeliger behavior compared to the Lambertian behavior of the surface. \newline
The Minnaert disk function is given by the following equation \citep{Minnaert_1941}:
\begin{equation}
    D_{M} = \mu_0^{k} \mu^{k-1}
\end{equation}
where $k$ is the Minnaert parameter, which depends on both the albedo and the phase angle.\newline
The Akimov disk function is given by \citep{Akimov_1988, Shkuratov_1994, Shkuratov_2003}:
\begin{equation}
    D_{A} =  \cos\left(\frac{\alpha}{2}\right) \left[\frac{\pi}{\pi - \alpha}\left( \gamma - \frac{\alpha}{2}\right)\right] \frac{\left(\cos \beta\right)^{\alpha/(\pi - \alpha)}}{\cos \gamma}
\end{equation}
The advantage of this version of the Akimov disk function is that it does not introduce free parameters. $\beta$ and $\gamma$ are the photometric latitude and the photometric longitude, respectively. These two variables are directly given by the illumination angles \citep{Shkuratov_2011}:
\begin{equation}
    \cos \beta = \sqrt{\frac{\sin^2(i+e) - \cos^2 \left(\frac{\varphi}{2}\right) \sin(2e) \sin(2i)} {\sin^2(i+e) - \cos^2 \left(\frac{\varphi}{2}\right) \sin(2e) \sin(2i) + \sin^2(e) \sin^2(i) \sin^2(\varphi) }}
\end{equation}
\begin{equation}
    \cos \gamma = \frac{\cos e}{\cos \beta}
\end{equation}
Therefore, we can define three different KS models:
\begin{equation}
    KS_1 = A_{N} \frac{e^{-\nu_{1} \alpha} + m e^{-\nu_{2}\alpha}}{1 + m} \left[c_{l} \frac{2 \mu_{0}}{\mu_{0}+\mu} + (1-c_{l})\mu_{0}\right]
\end{equation}
\begin{equation}
    KS_2 = A_{N} \frac{e^{-\nu_{1} \alpha} + m e^{-\nu_{2}\alpha}}{1 + m} \mu_0^{k} \mu^{k-1}
\end{equation}
\begin{multline}
    KS_3 = A_{N} \frac{e^{-\nu_{1} \alpha} + m e^{-\nu_{2}\alpha}}{1 + m} \\ \times \cos\left(\frac{\alpha}{2}\right) \left[\frac{\pi}{\pi - \alpha}\left( \gamma - \frac{\alpha}{2}\right)\right] \frac{\left(\cos \beta\right)^{\alpha/(\pi - \alpha)}}{\cos \gamma}
\end{multline}
The $KS_1$ and $KS_2$ model have both five free parameters ($A_N$, $\nu_1$, $\nu_2$, $m$, and $c_l$ or $k$). The $KS_3$ has four free parameters ($A_N$, $\nu_1$, $\nu_2$, $m$).

\section{Additional figures for disk-resolved photometry and spectro-photometry}
\begin{figure*}
\resizebox{\hsize}{!}{\includegraphics{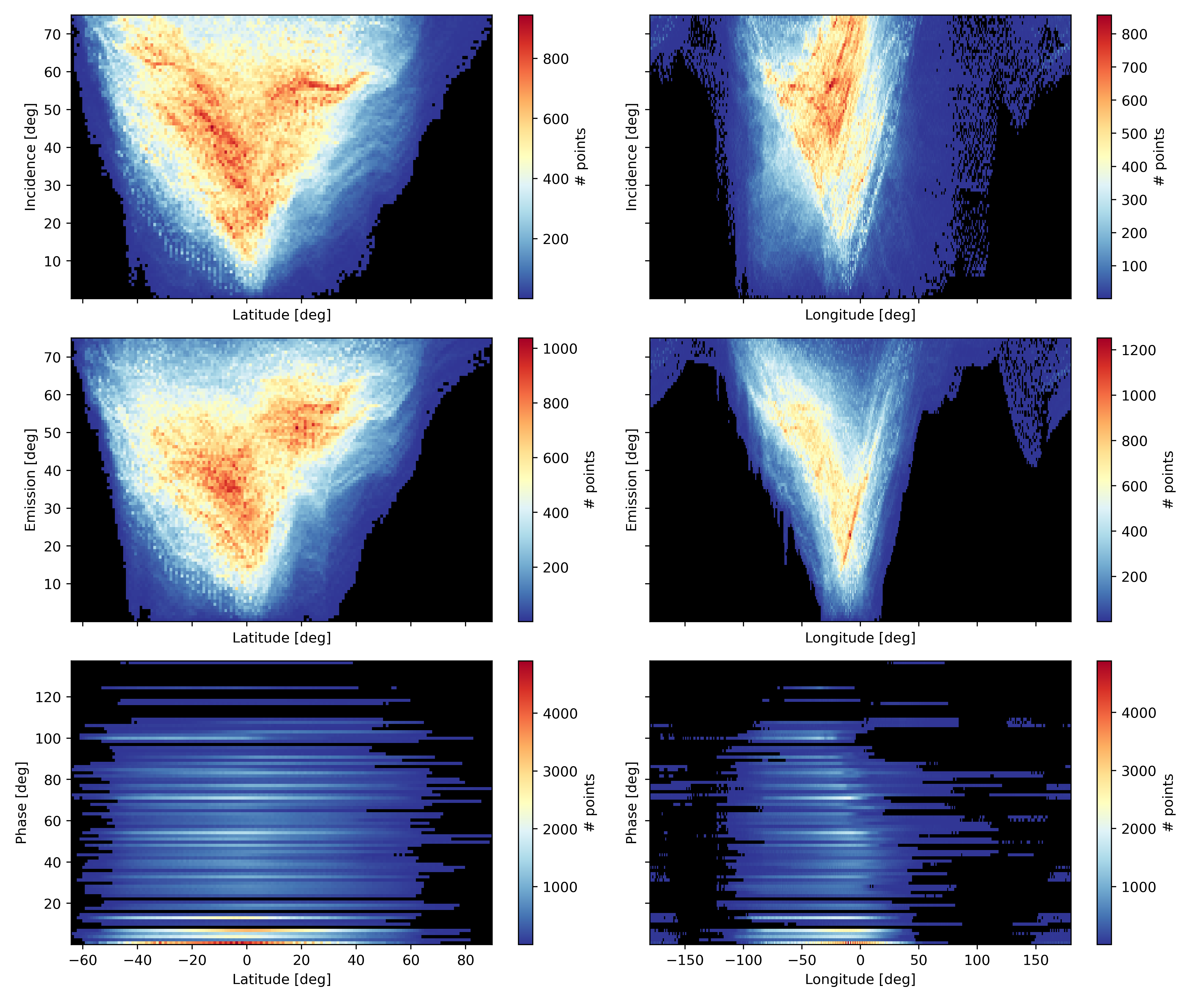}}
\caption{Density plots of incidence, emission, and phase angles coverage across Deimos surface with the SRC. Black areas represent regions with no data.}
\label{fig:SRC_ill_conditions}
\end{figure*}

\begin{figure*}
\resizebox{\hsize}{!}{\includegraphics{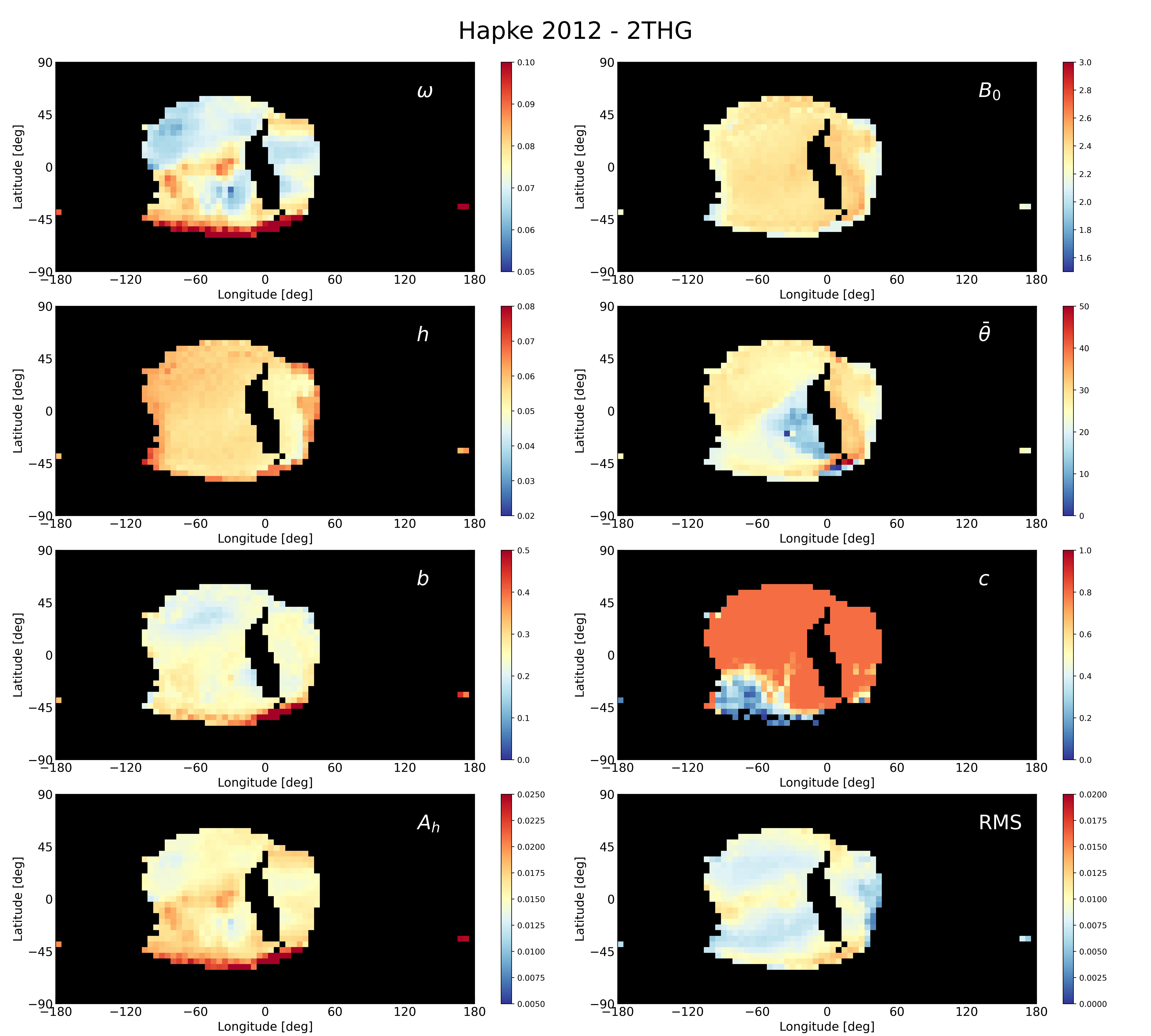}}
\caption{Hapke parameter maps (H2012-2THG) and hemispherical albedo $A_h$ derived with the SRC observations at 650 nm. The RMS error map appears to be linked with the position of the ridge. The data are projected on the map using the equirectangular projection.}
\label{fig:HRSC_params_map_2THG}
\end{figure*}

\begin{figure}
\resizebox{\hsize}{!}{\includegraphics{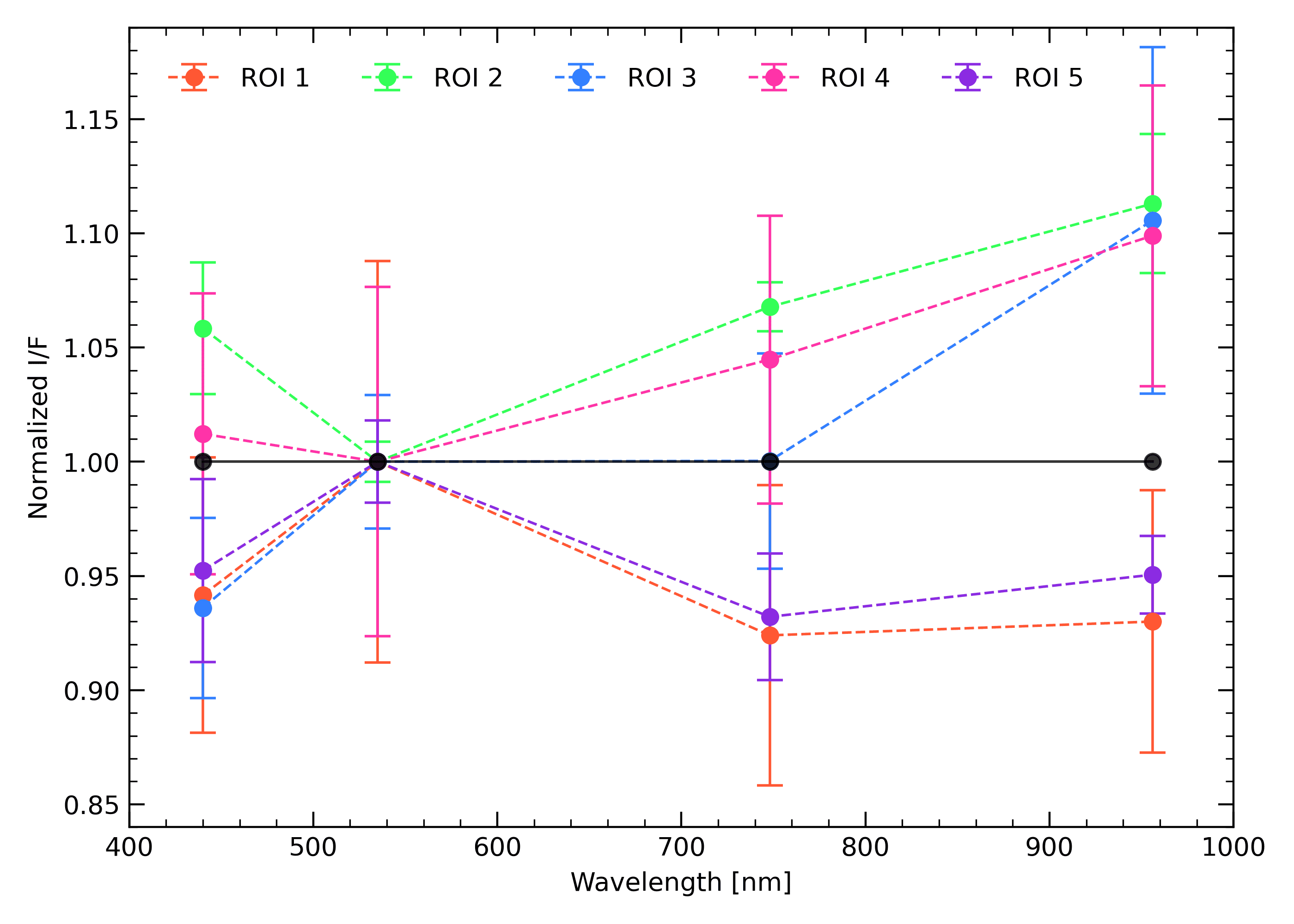}}
\caption{Ratio of the HRSC radiance factor in the four filters and the disk-integrated spectrum. All spectra are normalized at 535 nm.}
\label{fig:ratio_deimos_rois}
\end{figure}

This section presents several additional figures of the disk-resolved photometric analysis. Fig. \ref{fig:SRC_ill_conditions} shows the spatial coverage of the SRC dataset on Deimos, and the density of observations for the photometric analysis according to the incidence, emission, and phase angles. Fig. \ref{fig:HRSC_params_map_2THG} presents the Hapke parameter maps for the H2012-2THG model. Note that we did not succeed in fitting the $c$ parameter.\newline
Fig. \ref{fig:ratio_deimos_rois} shows the evolution of the radiance factor for the different filters from blue to IR in comparison with the disk-integrated spectrum. 

\section{Photometric correction}
\begin{figure*}
\resizebox{\hsize}{!}{\includegraphics{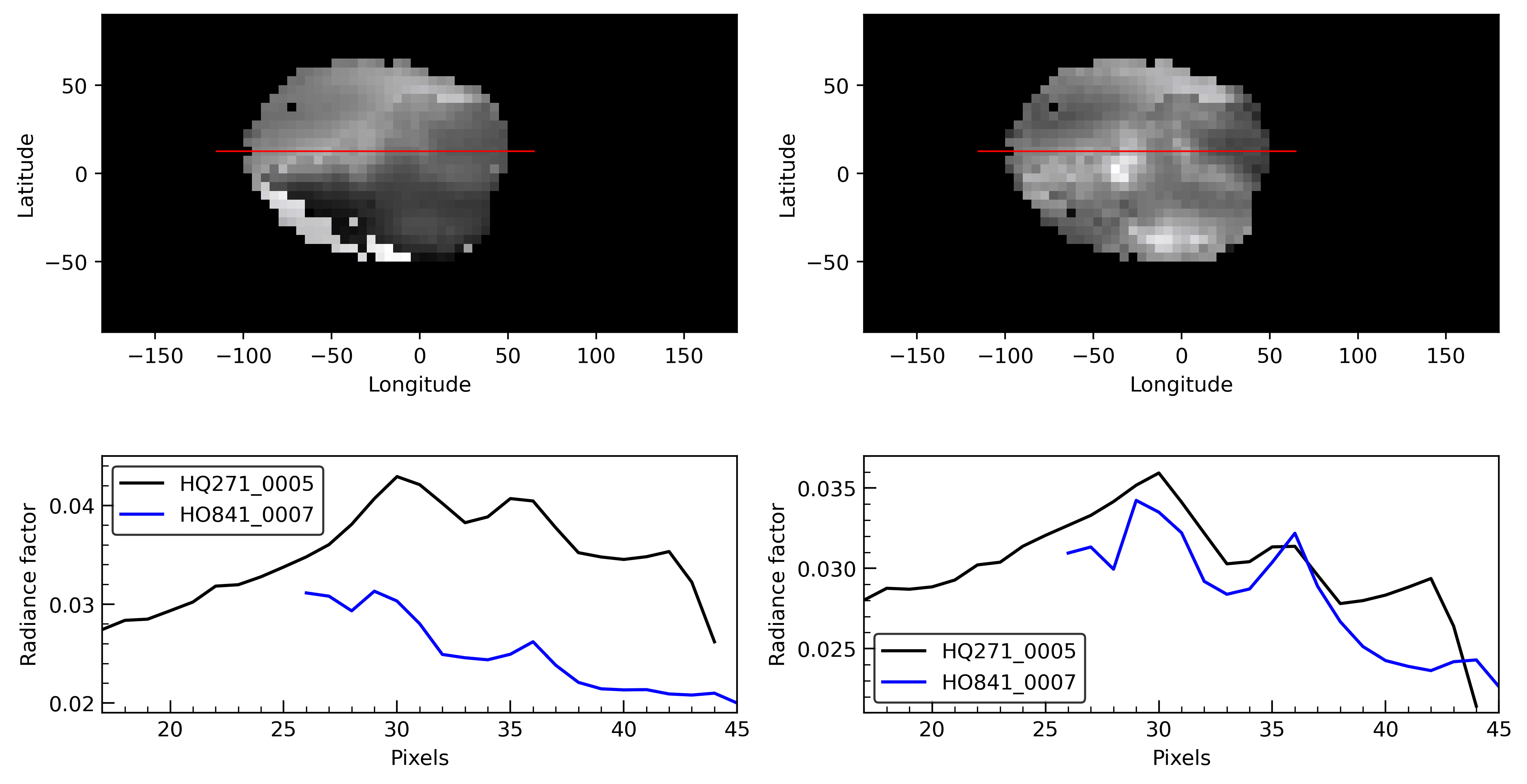}}
\caption{Mosaic map of Deimos using two images taken at different phase angle by the SRC camera. Image hq271\_0005 was acquired with a phase angle of 14.9° and the image ho841\_0007 with a phase angle of 40.6°. (\textit{Top left}) Mosaic of the two images without photometric correction. (\textit{Bottom left}) Profile of the reflectance for the two images along the red line plotted in the mosaic above. (\textit{Top and bottom right}) Same as previously but here with the McEwen photometric correction.}
\label{fig:photometric_correction_1}
\end{figure*}

\begin{figure}
\resizebox{\hsize}{!}{\includegraphics{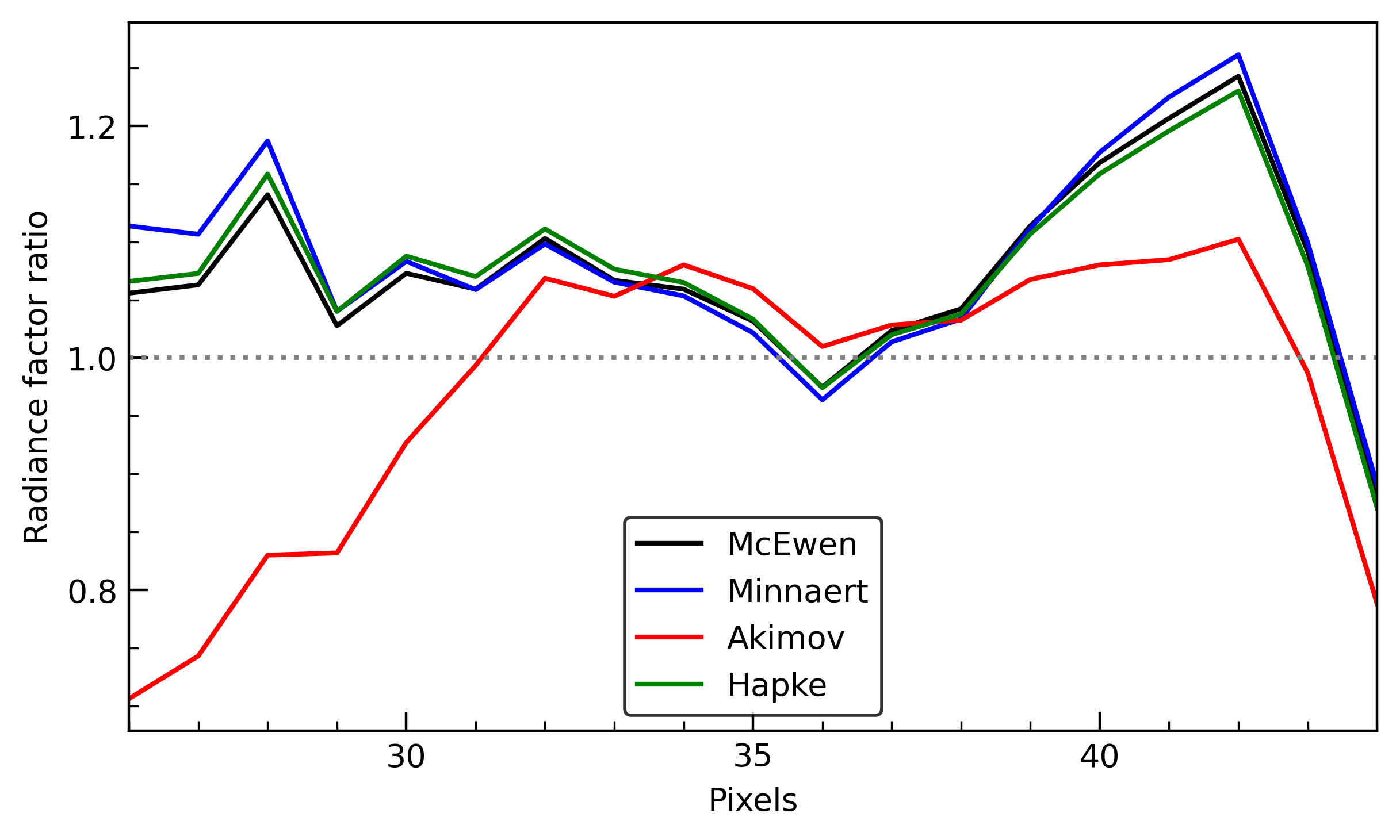}}
\caption{Radiance factor ratio of the plotted profiles from Fig. \ref{fig:photometric_correction_1}.}
\label{fig:photometric_correction_2}
\end{figure}

The derived photometric parameters from the global disk-resolved analysis of Deimos SRC data were also used to perform the photometric correction of the data. Fig. \ref{fig:photometric_correction_1} presents an example of the application of the McEwen photometric correction on a set of two images taken at different phase angles. Fig. \ref{fig:photometric_correction_2} shows the results of the photometric correction using the different models defined in this work: McEwen, Minnaert, Akimov, and Hapke.

\end{appendix}

\end{document}